\documentclass[]{article}
\usepackage[dvips]{graphicx}
\usepackage{amsfonts}
\usepackage{amsmath}
\usepackage{amssymb}

 \def\2{I$\!$I}

\usepackage{color}

\newcommand{\rd}{\textcolor{red}}

\def\*{{\phantom *}}
\begin{document}
\font\fifteen=cmbx10 at 15pt \font\twelve=cmbx10 at 12pt

\begin{titlepage}

\begin{center}
\renewcommand{\thefootnote}{\fnsymbol{footnote}}

\vspace{6 cm}

\hspace{6cm} D\'{e}di\'{e} \`{a}  Iridii Aleksandrovich  Kvasnikov, qui m'a appris\\
\hspace{8cm} \`{a} admirer la condensation de Bose-Einstein...

\vspace{2 cm}

{\fifteen A century of the Bose-Einstein condensation concept\\
\vspace{0.2 cm}
and \\
\vspace{0.2 cm}
half a century of the JINR experiments for observation \\
of condensate in the superfluid $^{4}$He (He II)}

\vspace{0.3cm}


\vspace{1.0cm}

{\rd{(juillet-ao\^{u}t-septembre 2025)}}

\vspace{1cm}

\setcounter{footnote}{0}
\renewcommand{\thefootnote}{\arabic{footnote}}

{\bf Valentin A. ZAGREBNOV}
\footnote{Institut de Math\'{e}matiques de Marseille - CNRS,
Universit\'{e} d'Aix-Marseille,
Campus Saint Charles, \\ 3 place Victor Hugo, Case 19,
13331 Marseille - Cedex 3, France
(courriel: Valentin.Zagrebnov@univ-amu.fr)}

\medskip

Institut de Math\'{e}matiques de Marseille - CNRS,\\
Universit\'{e} d'Aix-Marseille,
Campus Saint Charles, \\ 3 place Victor Hugo, Case 19,
13331 Marseille - Cedex 3, France


\vspace{0.5 cm}

{\bf Abstract}

\end{center}
This short review is devoted to celebration of two major events in quantum physics. 
The first one is the birth of the concept of Bose-Einstein condensation (1925) and the 
second is the experimental proof that it does exist and appears in the liquid  $^{4}$He 
\textit{simultaneously} with superfluidity below the $\lambda$-point (1975).

The both of these events are tightly related to the Bogoliubov theory of superfluidity (1947). 
The existence of condensate in the system of interacting bosons is the key \textit{ansatz} of 
this theory. Therefore, the experiments started in JINR in 1975 confirmed this prediction of the 
Bogoliubov theory that superfluidity of the liquid  $^{4}$He (He II) should emerge at the same 
time as the Bose-Einstein condensation.


\vspace{1cm}

\noindent \textbf{Keywords} : Bose-Einstein statistics and condensation,
conventional and non-conventional condensations, generalised condensations,
Bogoliubov theory of superfluidity, deep-inelastic neutron scattering,
Bose-Einstein condensate in liquid $^{4}$He, JINR-Dubna

\vspace{0.5cm}

\noindent \textbf{PACS} : 05.30.Jp, 02.70.Lq, 03.75.Fi, 67.40.D6

\end{titlepage}

\tableofcontents

\newpage

\vspace{-1.0cm}

\begin{flushright}
\parbox{7.5cm}{``Autant les physiciens sont inventifs dans leur capacit\'{e} \`{a} poser des questions
sur la Nature et \`{a} l'expliquer, autant les math\'{e}maticiens sont tout aussi inventifs pour
expliquer pourquoi les physiciens font les choses correctement ... .''
\newline

``Ich behaupte aber, dass in jeder besonderen Naturlehre nur so viel
eigentliche Wissenschaft angetroffen werden kann, als darin Mathematik anzutreffen ist.''\\
\textbf{Immanuel Kant,}\textit{\ Metaphysische Grundlagen der Naturwissenschaft, 1786.}
}
\end{flushright}

\vspace{1cm}

\section{Bosons and Bose-Einstein Condensation (1925)\label{section 1}}


\subsection{Many-body quantum systems and  the Bose-statistics}\label{subsection 1.1}

\subsubsection{Planck formula}\label{subsubsection 1.1.1}

The one hundred years ago predicted by A.Einstein \cite{Einstein} \textit{condensation}
of the Bose-gas (1925) went through more than decade of a strong doubt before it
has been widely accepted after a convincing elucidation and formal mathematical arguments by F.London
\cite{London} in 1938.

Although at that time even the Quantum Mechanics of {few particles} has not yet been
completely formulated (the Schr\"{o}dinger equation was published only in 1926 \cite{Schrodinger})
physicists were already puzzled by the problem of describing the conductivity, or specific heat,
of \textit{many-body} quantum systems such as electrons in metals. Another problem of this kind
concerned a formula for the spectral density of electromagnetic radiation emitted by a black body
in thermal equilibrium at given temperature $T$. Once M.Planck \cite{Planck} had discovered for
it an empirically fitting law, the \textit{Planck formula} (1900), it was necessary to find
satisfactory (mathematical) arguments which yield a derivation of this formula.

Planck's arguments were formal and they were based on two hypotheses : \\
- The \textit{first} one was to attribute to any \textit{infinitesimal} band $d\nu$
(for \textit{mode} $\nu$) of radiation in a closed cavity at given temperature $T$ a (proportional to
\textit{volume} of the cavity) system of $N$ monochromatic vibrating \textit{resonators} with the
proper frequency $\nu$. \\
- The \textit{second} hypothesis was that the energy $\varepsilon_n$ of each resonator is 
\textit{quantified}
according to the law: $\varepsilon_n = n \, h\nu$.
Here $n = 0,1,2, \ldots $, and $h=6,6262\cdot10^{-27} \, erg\cdot sec$ is the \textit{Planck constant}.
Then Planck has named the product $h\nu$ by the \textit{energy element} and supposed that a given
\textit{exited} resonator may possess $n \geq 0$ of these elements.

As a consequence, any configuration $\mathcal{C}$  of the system of $N$ {exited/non-exited} resonators
will have a total energy $M_{\mathcal{C}} \, h\nu$ proportional to the \textit{energy element} $h\nu$,
which is also known as the (single) \textit{light-quantum} energy (A.Einstein (1905)\cite{Einstein05}).

Seeing that the cavity of the oven is \textit{thermal} and in order to establish contact with the
temperature of radiation, Planck appealed, in the next step, to Statistical Mechanics. To this aim 
he first
calculated the \textit{statistical weight} $\Gamma_N(M)$ of configurations in the system of $N$
resonators for a \textit{fixed} energy $E = M \, h\nu$, i.e., in the \textit{microcanonical} ensemble,
see for example, \cite{Kvasnikov}, pp.603-604.
Note that in fact the number $\Gamma_N(M)$ of distributions of $M$ {light-quanta} over $N$ resonators
coincides with the number of possibilities of distributing $M$ \textit{objects} into $N$ 
\textit{boxes}:
\begin{equation}\label{M-N}
\Gamma_N(M) = \frac{(M + N - 1)!}{M! (N - 1)!} \, .
\end{equation}
Applying the \textit{Stirling formula} for large $N$ and $M$ in (\ref{M-N}), Planck then
followed the \textit{Boltzmann principle} for deducing the asymptotic form of the microcanonical
\textit{entropy}
\begin{equation}\label{S-Pl}
S(M,N)):= \ln \Gamma_N(M) = (M + N)\ln (M + N) - M \ln M  - N \ln N \, ,
\end{equation}
and the corresponding to the system of $N$ resonators temperature 
$T$: $1/(k_B T) := \partial_E S(E/h\nu,N)$,
here $k_B = 1,38 \cdot 10^{-16}\, erg/K$ is the \textit{Boltzmann constant}. Then (\ref{S-Pl}) 
yields
\begin{equation}\label{kB-T}
\frac{1}{k_B T}= \partial_E \ln \Gamma_N(M) = \frac{1}{h\nu}\ln \left(1 + \frac{N}{M}\right) \, .
\end{equation}
Next Planck defined by $\varepsilon(\nu) = E/N$ the \textit{mean-value} energy for resonators in
the system of $N$ {exited/non-exited} resonators and deduced from (\ref{kB-T}) that
\begin{equation}\label{varepsilon-nu}
\varepsilon(\nu) = \frac{h\nu}{e^{\beta \, h\nu} - 1}  \, .
\end{equation}
Finally, multiplying (\ref{varepsilon-nu}) by
\textit{density-number} of radiation modes \textit{per} unit volume and \textit{per} infinitesimal band,
taken from the \textit{classical} electrodynamics he obtained
\begin{equation}\label{Spec-dens}
\rho(\nu,T) = \frac{8 \pi \nu^2}{c^3}\frac{h\nu}{e^{\beta \, h\nu} - 1} \, ,
\quad \beta = \frac{1}{k_B T} \, ,
\end{equation}
which is the \textit{spectral density} of radiation emitted by a black-body at a given temperature $T$,
that is, the \textit{Planck formula} (1900).

We note that Planck did not attribute any definite physical significance to his hypothesis of
\textit{resonators} but rather proposed it as a mathematical \textit{device} that enabled him to derive
an expression for the black-body spectrum that matched the empirical data for all frequencies
$\nu$. Although initially he was inclined to consider the \textit{light-quanta} of resonators only
for exchanging by the energy between radiation and the oven walls, at the end of the paper \cite{Planck}
Planck has tentatively mentioned the possible connection of such \textit{resonators} with a
mono-atomic gas.

Much later A.Einstein (1916) \cite{Einstein16} provided a different
demonstration of the Planck formula. His arguments involved the idea of the \textit{light-quanta} $h\nu$,
but also their \textit{interactions} with a gas of \textit{two-level molecules} (with the level spacing
according to the Bohr rule: $E_2 - E_1 = h\nu$) occupying the cavity, which has temperature $T$,
see, e.g., \cite{Kvasnikov}, pp.604-607, for details.

\subsubsection{S.N.Bose and Bose-Einstein statistics}\label{subsubsection 1.1.2}

Recall that A.Einstein (1905) had succeeded to explain the \textit{photoelectric emission} because
he supposed that the light "...\textit{consists of a finite number energy quanta are spatially 
localized
at points of space, move without dividing and are absorbed or generated only as a whole}"
\cite{Einstein05}. Moreover, the \textit{Compton effect} \cite{Compton} (1923) had also
clearly indicated that radiation \textit{itself} consists of \textit{light-quanta} (they were named
\textit{photons} by G.N.Lewis \cite{LewisGN} in 1926). But the mathematically satisfactory quantum
theory of the \textit{light} considered as a \textit{many-body} quantum system, with elucidation of the
black-body radiation and the Planck law, was developed by Satyendra Nath Bose \cite{Bose} only in 1924.

Originally Bose had submitted his manuscript in the \textit{Philosophical Magazine} (Taylor \& Francis),
but it was rejected there. He then sent it to A.Einstein with the humble request: "... If you think
the paper worth publication, I shall be grateful if you arrange its publication in Zeitschrift f\"{u}r
Physik". In a footnote to the translated into German paper, Einstein wrote: "In my opinion Bose's 
derivation
signifies an important advance. The method used here gives the quantum theory of the ideal gas
(that is, of atoms, or molecules - \textit{remark by VAZ}), as I will explain elsewhere."

In his pioneering paper \cite{Bose} Bose \textit{extended} Planck's method of quantisation of "imaginary"
vibrating resonators, which by energy quanta directs the connection between radiation and matter
(the oven walls, or Planck's "speck of carbon" \cite{Planck14}),
to the \textit{quantisation} of the electromagnetic field in cavity, which implies a
non-classical \textit{corpuscular} nature of radiation \textit{itself} in the spirit of the
{light-quanta} \cite{Einstein05} !

To this aim he started in \cite{Bose} by a \textit{plain-spoken} declaration that \textit{light-quantum}
(photon) with energy $\varepsilon = h\nu$ has a \textit{momentum} $p = \sqrt{p_x^2 + p_y^2 + p_z^2}$ of
the magnitude $p = h\nu/c$ in \textit{direction} of its movement. \\
Next, Bose proceeded with calculation in cavity a \textit{density of states} for
photon-momentum {operator} $\widehat{p} = (\hbar/i) \nabla$, where $\hbar = h/2\pi$.
For a cubic cavity: $\Lambda = L \times L \times L \subset \mathbb{R}^3$,
with \textit{periodic boundary} conditions one gets explicitly the eigenfunctions
$\psi_{\textbf{k}}(\textbf{x}) = \exp(i \, \textbf{k}\cdot \textbf{x})$
and the spectrum $\sigma(\widehat{p})$ of the self-adjoint operator $\widehat{p}$ in a
standard for the quantum mechanics Hilbert space $\mathcal{H} = L^2(\Lambda)$ of the square-integrable
complex wave-functions in $\Lambda \subset \mathbb{R}^3$. They are enumerated
by wave-vectors $\textbf{k} = (k_x, k_y, k_z)\in \sigma(\widehat{p})/\hbar$:
\begin{equation}\label{Eigenfun-Spec}
\widehat{p}\, \psi_{\textbf{k}} = \hbar \, \textbf{k} \, \psi_{\textbf{k}} \ \  {\rm{and}} \ \
\textbf{k} \in \{2\pi/L \, (s_x, s_y, s_z): \, (s_x, s_y, s_z) \in \mathbb{Z}^3\} =: \Lambda^*
 \, .
\end{equation}
Then number of the photon states in the infinitesimal band $[\nu, \nu + d\nu]$ is the number $dN$ of
eigenvectors in the volume of corresponding spherical shell $[k, k+dk]$ divided by the
\textit{$k$-lattice volume} per point (\ref{Eigenfun-Spec}). Since $p = \hbar k = h \nu /c  $,
one gets
\begin{equation}\label{dN}
dN(k) = \frac{4 \pi k^2 dk}{(2\pi/L)^3} = L^3 \  \frac{4 \pi \nu^2 d\nu}{c^3} \, .
\end{equation}
Now taking into account \textit{two} states of \textit{polarisation} of light (the \textit{photon spin}
orientations either parallel, or antiparallel to its direction of motion) Bose deduced from
(\ref{dN}) for density of the photon states (modes), per unit volume of cavity and per infinitesimal
band, the expression
\begin{equation}\label{density}
J(\nu) := \frac{8 \pi \nu^2}{c^3} \, .
\end{equation}
The value of (\ref{density}) coincides with the first factor in the
Planck formula (\ref{Spec-dens}) but with a new meaning: it is the one-particle photon density
of states in the mode $\nu$ for the unit volume of cavity.

Finally, Bose considered photons $h\nu$ as identical \textit{indistinguishable} particles allowed
(similar to the Planck \textit{energy elements}) that \textit{finitely} many of them may accumulate
in a single photon state with wave-vector $\textbf{k}$. Therefore, any configuration of photons in 
cavity can
be labelled by \textit{occupation numbers} $\{N_\textbf{k} = 0,1,2, \ldots\}_{\textbf{k}\in \Lambda^*}$.
With this \textit{prescription} for counting the photon configurations in a hot cavity with  photon
density of states (\ref{density}) and at the temperature $T$, Bose proved in \cite{Bose}
the Planck formula (\ref{Spec-dens}) for spectral density of the black-body radiation emitted by cavity.
To this end (similarly to M.Planck) he applied in the last part of \cite{Bose} the entropy variational
principle of Statistical Mechanics for states in equilibrium, cf. \cite{Kvasnikov} Ch.II, Problems \S4.

Bose's very natural \textit{ansatz} about \textit{indistinguishability} of photons ("bosons")
turned out to be far-reaching. His \textit{receipt} for counting the allowed configurations of many-body
photon system was then extended by Einstein \cite{Einstein24} to the
\textit{mono-atomic ideal} quantum gas. Here instead of momentum operator 
$\widehat{p} = (\hbar/i) \nabla$
(\ref{Eigenfun-Spec}) one considers for the gas of atoms with mass $m$ in box $\Lambda$ the
one-particle \textit{kinetic-energy} operator $T_{\Lambda} = \widehat{p}^{\,2}/2m$.
For periodic boundary conditions on $\partial\Lambda$ it has eigenvalues/eigenfunctions:
\begin{equation}\label{Eigenfunc-Spec-m}
T_{\Lambda} \, \psi_{\textbf{k}} = \varepsilon_\textbf{k} \psi_{\textbf{k}} \ \  {\rm{and}} \ \
\textbf{k} \in \{2\pi/L (s_x, s_y, s_z): \, (s_x, s_y, s_z) \in \mathbb{Z}^3\} = \Lambda^*
 \, ,
\end{equation}
with eigenvalues $\varepsilon_\textbf{k} := (\hbar \, \textbf{k})^2/2m \, $ for eigenfunctions
$\{\psi_{\textbf{k}}(x) = \exp(i \, \textbf{k}\cdot \textbf{x})\}_{\textbf{k} \in \Lambda^*}$.
This passage from photons to the
quantum ("boson") gas of atoms evidently modifies in $\Lambda$ the one-particle density of states,
but not the prescription (\textit{statistics}) for counting of allowed configurations of many-body
particle system since by virtue of the \textit{indistinguishability} they still can be labelled only
by \textit{occupation numbers} $\{N_\textbf{k}\}_{\textbf{k}\in \Lambda^*}$ of the one-body 
vector states
$\{\psi_{\textbf{k}}(x)\}_{\textbf{k} \in \Lambda^*}$.

Note that owing to the Schr\"{o}dinger equation for the ideal quantum gas in $\Lambda$, the $N$-particle
eigenfunctions $\Psi_{l}^{(N)} (x_1, \ldots, x_N) =
\Psi_{\{\textbf{k}_{j}: 1\leq j \leq N\}}^{(N)} (x_1, \ldots, x_N)$, for $N$-particle operator
$T_{\Lambda}^{(N)} := \sum_{1\leq j\leq N} \widehat{p_j}^{\,2}/2m$, can be presented
as linear combinations of products:
\begin{equation}\label{Eigenfunc-N}
\prod_{j=1}^N \psi_{\textbf{k}_j}(x_j) =
\prod_{\textbf{k}\in \Lambda^*} \prod_{j(k)=1}^{N_\textbf{k}} \psi_{\textbf{k}}(x_{j(\textbf{k})}) =
\Phi_{\{N_\textbf{k}\}_{\textbf{k} \in \Lambda^*}}^{(N)} (x_1, \ldots, x_N)\, .
\end{equation}
Hence, the occupation number $N_\textbf{k}$ coincides with with the number of \textit{identical}
wave-functions $\{\psi_{\textbf{k}}(x_{j(\textbf{k})})\}$ in the left-hand side product of identity
(\ref{Eigenfunc-N}).
This occupation number has upper bound: $N_\textbf{k} \leq N$, and for $N_\textbf{k} = N$ the
\textit{symmetric} function $\prod_{1\leq j \leq N}\psi_{\textbf{k}}(x_j)$ in (\ref{Eigenfunc-N})
describes $N$ \textit{indistinguishable} particles in the box $\Lambda$ occupied a single mode with
$\textbf{k}\in \Lambda^*$. To keep
\textit{indistinguishability} for general configurations of {occupation numbers}
$\{N_\textbf{k}\}_{\textbf{k}\in \Lambda^*}$ one has to \textit{symmetrise} the functions in
(\ref{Eigenfunc-N}). Therefore, the idea of the \textit{boson} quantum systems, or
\textit{Bose-Einstein statistics} was born in 1924 thanks to the papers \cite{Bose} and 
\cite{Einstein24}.

Now we exploit this idea to elucidate that the concept of \textit{indistinguishability} of $N$
identical quantum particles has an important consequence due to the quasi-classical \textit{thought}
experiments concerning a permutation, for example, of {two} of them. In fact, this experiment implies
that being a unique solution of the Schr\"{o}diger equation the \textit{normalised} wave-function for
indistinguishable particle stays (up to a phase factor $e^{i \phi}$) invariant with respect to
permutation of the corresponding to the arguments of these particles :
\begin{equation}\label{permutation}
\Psi_N (x_1, \ldots, x_s,\ldots, x_r,\ldots, x_N) = e^{i \phi} \,
\Psi_N (x_1, \ldots, x_r,\ldots, x_s,\ldots, x_N)   \, .
\end{equation}
Then because of (\ref{permutation}) one gets after second permutation that $e^{2i \phi} =1$,
and consequently $e^{i \phi} = \pm 1$.
For this reason the wave-functions of identical \textit{indistinguishable} quantum particles
can be only of two categories: \textit{symmetric}, corresponding to the \textit{Bose-Einstein} statistics
(1924) for \textit{bosons}, or \textit{antisymmetric}, corresponding to the \textit{Fermi-Dirac} statistics
(1926) for \textit{fermions}.

These ("non-local") \textit{collective} properties of the quantum statistics for
non-interacting \textit{indistinguishable} particles conflicts fundamentally with the
\textit{Boltzmann statistics}, which ensures a \textit{statistical independence} of non-interacting
classical particles. We clarify that in the case of (Bose-Einstein, or Fermi-Dirac)
quantum statistics for \textit{indistinguishable} particles a list of occupation numbers
$\{N_\textbf{k}\}_{\textbf{k}\in \Lambda^*}$ defines one and \textit{unique} state
$\Psi^{(N)}_{\{N_\textbf{k}\}_{\textbf{k}\in \Lambda^*}}$ of the quantum system. Whereas, if
particles are \textit{classical}, one has to \textit{enumerate} them. Then besides the list
$\{N_\textbf{k}\}_{\textbf{k}\in \Lambda^*}$ their classical \textit{microstates} ({configurations})
depend on distribution of attributed \textit{labels}. For the partition function this produces a 
supplementary
degeneracy factor $N!/\prod_{k} N_\textbf{k}!$, which yields the \textit{Boltzmann counting} of allowed
configurations and thus the \textit{quasi-classical} limit of quantum system corresponding to
\textit{rarified} classical ideal gases. For details see, for example, \cite{Kvasnikov}, Ch.III, \S 1.

Next, by analysis of \textit{correlations} we strengthen the key observation that in contrast to the
\textit{statistical independence} (in the standard of this term) of particules of the classical 
ideal gas,
the particles of an ideal quantum gas are \textit{indistinguishable}, but \textit{not} independent.
For this one can consider \textit{two-point correlation functions} for (\textit{spinless})
bosons $F_{B}(R)$ and fermions $F_{F}(R)$ separated by inter-particle distance
$R = |\textbf{r}_1 -\textbf{r}_2|$.
Then explicit calculations (see \cite{Kvasnikov4}, Ch.1, \S 4 \textbf{e})) for $N$-body ideal
Bose $(+)$ and Fermi $(-)$ gases in volume $V = |\Lambda|$ yield
\begin{equation}\label{two-point-BF}
F_{B/F}(R) = 1 \pm v^2 \left|\frac{1}{V}\sum_{\textbf{k} \in \Lambda^*}
n_{k}^{B/F} \, e^{i \, \textbf{k}\cdot\textbf{R}}\right|^2
\, ,  \quad v=V/N \, ,
\end{equation}
Here the \textit{grand-canonical} Gibbs expectation (\textit{mean-value}) of the occupation number $N_k$
for bosons is: $n_{k}^{B} = (e^{\beta (\varepsilon_k -\mu)} -1)^{-1}$, whereas for fermions it is:
$n_{k}^{F} = (e^{\beta (\varepsilon_k -\mu)} +1)^{-1}$, and $\mu$ is the value of the chemical potential
in the grand-canonical ensemble. Then \textit{out} from the quantum \textit{degenerate} regime
(for small densities, high temperatures $\theta = {k_B T}$ , that is, when the \textit{thermal}
de Broglie wave-lengths: $\lambda_{deB}(\theta) = \hbar/\sqrt{m \, \theta}$, is much \textit{smaller}
than the average \textit{inter-particle distance} $\sqrt[3]{v}$), we obtain from (\ref{two-point-BF})
that, already within the first quantum corrections to the case of the \textit{Boltzmann statistics}:
$F(R) = 1$, the quantum correlations do exist:
\begin{equation}\label{two-point-BF-nonDeg}
F_{B/F}(R) \simeq 1 \pm e^{- (m \theta /\hbar^2)\, R^2} \, ,
\quad \hbar/\sqrt{m \, \theta} \ll \sqrt[3]{v} \, .
\end{equation}
The interpretation of formula (\ref{two-point-BF-nonDeg}) is straightforward:\\
- Since for finite distances the correlation $F_{B}(R) > 1 $, the bosons be affected by a
\textit{temperature dependant} statistical \textit{attraction} to each other. It is monotonously
decreasing to non-correlated for ideal gas in the \textit{Boltzmann regime}: $F(R) = 1$, for the growing
$R$.\\
- Since for finite distances the correlation $F_{F}(R) < 1 $, the fermions be affected by a
\textit{temperature dependant} statistical \textit{repulsion} from each other with the same as for bosons
behaviour for the growing $R$, but (to bolster the \textit{Pauli exclusion principle}) with a strong
\textit{repulsion} for $R\downarrow0$.\\
In paper \cite{Uhl-Gro32} the authors transformed these quantum \textit{temperature dependant}
statistical correlations into particle \textit{two-body potential} to treat the problem classically.
In fact, the classical (i.e., the \textit{Boltzmann}) limit of (\ref{two-point-BF}) (and thus
(\ref{two-point-BF-nonDeg})) corresponds to the high-temperature limit $\theta \rightarrow \infty$.

Note that in both cases (\ref{two-point-BF-nonDeg}) the \textit{effective radius} of correlation
$R_{corr}$ has order of the \textit{thermal} de Broglie wave-lengths:
$R_{corr} \sim \lambda_{deB}(\theta)$. For the quantum \textit{degenerate} regime that is,
when the \textit{thermal} de Broglie wave-lengths $\lambda_{deB}$ is comparable, or larger
than the average \textit{inter-particle distance} $\sqrt[3]{v}$), the calculations
yield for (\ref{two-point-BF}) qualitatively the same $R$-behaviour as for (\ref{two-point-BF-nonDeg}),
but with a larger \textit{effective radius} of correlations. A peculiarity, that concerns the Bose-gas,
is the limit (see \cite{Kvasnikov4}, Ch.1, \S 4 \textbf{e})):
\begin{equation}\label{two-point-BEC}
\lim_{R\rightarrow 0} F_{B}(R) = 2 -
\left(\frac{\langle N_{0}\rangle(\theta)}{N}\right)^2 \, , \quad \hbar/\sqrt{m \,\theta} 
\geq \sqrt[3]{v}\ ,
\end{equation}
where $\langle N_{0}\rangle(\theta)$ is the Gibbs expectations (\textit{mean-value})
of the occupation number in the mode $\textbf{k}=0$, see (\ref{Eigenfunc-Spec-m}). If this value is
\textit{macroscopic}, that is:
$\lim_{N\rightarrow\infty} \langle N_{0}\rangle(\theta)/N = {v} \, \rho_0(\beta) > 0$ for 
$\beta =1/\theta $,
then $\rho_0(\beta)$ is density of the \textit{Bose-Einstein condensate},
which will be a main subject of the next section.

\subsection{Bose-Einstein condensation
\label{subsection 1.2}}

\subsubsection{Conventional Bose-Einstein condensation and  the G.E.Uhlenbeck objections}
\label{subsubsection 1.2.1}

Based on \cite{Einstein24}(1924), Einstein applied then in \cite{Einstein}(1925)
the ideas of Bose-Einstein statistics to study the thermodynamic properties of the 
\textit{Ideal Bose-gas}
(IBG) and predicted in this system a peculiar \textit{condensation} of particles.
This phenomenon occurs in quantum \textit{degenerate} regime and manifests as a \textit{macroscopic}
(proportional to the volume of the system) mean-value of occupation number in \textit{one} of the modes.
But two years later G.E.Uhlenbeck in his doctoral thesis "On statistical methods in the quantum theory"
(Leyden, 1927) \cite{Uhlenbeck} criticised Einstein's arguments in favour of condensation on the
mathematical
ground. In particular, his critical remarks concern: the quantisation in finite volume, the 
implementation
of the thermodynamic limit and the accuracy with the replacement of certain sums by integrals,
see \cite{Uhlenbeck}, pp.69-71. We return to details of the Uhlenbeck objections below.

This criticism delayed the general acceptance of this \textit{conventional} one-mode
\textit{Bose-Einstein condensation} (BEC) for almost a decade. It is discovery of \textit{superfluidity}
of the liquid helium $^{4}\rm{He}$ with $\lambda-$point phase transition at $T = 2,172$ K
(for pressure $1$\textit{atm}) by P.L.Kapitza (1938)\cite{Kapitza} and J.F. Allen, A.D. Misener (1938)
\cite{Allen} that renewed interest to the BEC.
For example, in \cite{London} (1938) F. London wrote: "In discussing some properties of liquid helium,
I recently realized that Einstein's statement has been erroneously discredited; moreover, some support
could be given to the idea that the peculiar phase transition ("$\lambda-$point") ... very probably has
to be regarded as the condensation phenomenon of the \textit{Bose-Einstein statistics}, ... ."

The paper \cite{London} answered to the objections formulated in \cite{Uhlenbeck}, pp.69-71, 
and elucidated
the formal mathematical origin of the {conventional} one-mode BEC. In paper \cite{KahnUhlenbeck}
Uhlenbeck withdrew his objections. In fact, the arguments presented by F.London were similar to
the modern "finite-size scaling" approach to analysis of the phase transitions. To proceed, 
we demonstrate
below his line of reasoning, which is now widely accepted. Subsequently it also provided a
\textit{generalisation} of the conventional {concept} of the BEC for ideal and interacting boson
systems, see comments in subsection \ref{subsubsection 1.2.2} (see \cite{Berg1}, \cite{BergLewisPule2})
and in subsection \ref{subsubsection 1.2.3} (cf. \cite{BergLewisPule1}, \cite{BergDorlasLewisPule1},
\cite{ZagBru01}).

\vspace{0.2cm}
To this aim let $N$-particle ideal Bose-gas be enclosed in the \textit{cubic} box
$\Lambda =L\times L\times L \subset \mathbb{R}^3$,
$\left|\Lambda \right| =V$. We consider a possibility of the conventional BEC of the IBG in the
thermodynamic limit $V \rightarrow \infty$.
To this end (similarly to subsection \ref{subsubsection 1.1.2}) we consider in Hilbert space
$\mathcal{H} = L^2(\Lambda)$ the self-adjoint extension  of the one-particle kinetic-energy Hamiltonian
\begin{equation}\label{1-part-Ham}
T_{\Lambda}:=\left( -\frac{\hbar^2}{2 m}\Delta \right)_{\Lambda,{{p.b.}}} ,
\end{equation}
with \textit{periodic boundary} (p.b.) conditions  on $\partial \Lambda$. Then the one-particle spectrum
is $\sigma (t_{\Lambda}):=\{\varepsilon_\textbf{k} :=
\hbar^2 \textbf{k}^2/(2m)\}_{\textbf{k}\in\Lambda ^{\ast}}$, where $m$
denotes the mass of the particle and
\begin{equation}\label{k-Lambda}
\Lambda^{\ast }:= \{k_{j}= 2\pi n_{j}/L : n_{j} \in \mathbb{Z} \}_{j=1}^{3}
\end{equation}
is a \textit{dual} to $\Lambda$ (with respect to p.b.conditions) set of wave vectors.
Recall that by virtue of (\ref{Eigenfunc-Spec-m}), (\ref{Eigenfunc-N}) and due to the Bose-Einstein
statistics, the symmetric \textit{eigenfunctions} $\{\Psi_{l}^{(N)}\}_{l\geq 1}$
of the self-adjoint kinetic-energy Hamiltonian $T_{\Lambda}^{(N)}$ for $N$-particle ideal Bose-gas,
are entirely and uniquely determined  by \textit{configurations} of occupation numbers
$\{N_\textbf{k}\geq 0\}_{\textbf{k} \in \Lambda^*}$ in modes $\textbf{k} \in \Lambda^*$ for corresponding
functions $\Psi_{\{N_\textbf{k}\}_{\textbf{k} \in \Lambda^*}}^{(N)}$, such that
\begin{equation}\label{Eigenfunc-Spec-N}
T_{\Lambda}^{(N)} \Psi_{\{N_\textbf{k}\}_{\textbf{k} \in \Lambda^*}}^{(N)} =
\sum_{\textbf{k} \in \Lambda^*}\varepsilon_\textbf{k} N_\textbf{k} \,
\Psi_{\{N_\textbf{k}\}_{\textbf{k} \in \Lambda^*}}^{(N)}.
\end{equation}
This bijection reduces the quantum Gibbs calculations for \textit{canonical ensemble} $(\theta, V, N)$
to Statistical Mechanics on \textit{configurations} of occupation numbers (cf. (\ref{Eigenfunc-Spec-N}))
for \textit{canonical} partition function:
\begin{equation}\label{canon-ens}
Z_{\Lambda}(\theta,V,N) = {\rm{Tr}}_{\mathcal{H}^N} \, (e^{- \beta \, T_{\Lambda}^{(N)}}) =
\sum_{\{N_\textbf{k}\geq 0\}_{\textbf{k} \in \Lambda^*}}
e^{- \beta \sum_{\textbf{k} \in \Lambda^*} \varepsilon_\textbf{k} N_\textbf{k}} \
\delta (N, \sum_{\textbf{k} \in \Lambda^*} N_\textbf{k}) \, , \quad \beta = 1/\theta \, .
\end{equation}
Here $\delta(x,y)$ for $x,y \in \mathbb{N}_0$ is the \textit{Kronecker} symbol. To escape this constraint
and to profit of explicit calculations (as in  \cite{London}) we pass to the
\textit{grand-canonical ensemble} $(\theta, V, \mu)$, where the chemical potential $\mu < 0$ controls
\textit{total} number of particles in $\Lambda$. Then, the \textit{grand-canonical} partition function
gets the form:
\begin{equation}\label{grand-canon-ens}
\Xi_{\Lambda}(\theta,V,\mu) = \sum_{N = 0}^\infty e^{\beta \mu \, N} Z_{\Lambda}(\theta,V,N) =
\sum_{\{N_\textbf{k}\geq 0\}_{\textbf{k} \in \Lambda^*}} \prod_{\textbf{k} \in \Lambda^*}
e^{- \beta (\varepsilon_\textbf{k} - \mu) N_\textbf{k}} =
\prod_{\textbf{k} \in \Lambda^*} \sum_{N_\textbf{k} = 0}^\infty
e^{- \beta (\varepsilon_\textbf{k} - \mu) N_\textbf{k}} .
\end{equation}
As a consequence of (\ref{grand-canon-ens}), the \textit{grand-canonical} Gibbs probability
distribution is a \textit{product-measure} such that for expectations (\textit{mean-values})
$n_q (\beta,\mu ):= \left\langle N_{q}\right\rangle_{T_{\Lambda }}(\beta,\mu )$
of occupation numbers $N_{\textbf{q}}$ in any mode $\textbf{q}\in\Lambda^{\ast}$ one obtains:
\begin{equation}\label{q-dens}
\left\langle N_{q}\right\rangle_{T_{\Lambda }}(\beta,\mu ) :=
\frac{1}{\Xi_{\Lambda}(\theta,V,\mu)} \prod_{\textbf{k} \in \Lambda^*\setminus q}
\sum_{N_\textbf{k} = 0}^\infty e^{- \beta (\varepsilon_\textbf{k} - \mu) N_\textbf{k}}
\sum_{N_\textbf{q} = 0}^\infty e^{- \beta (\varepsilon_\textbf{\textbf{q}} - \mu)
N_\textbf{\textbf{q}}} N_{\textbf{q}} =
\frac{1}{e^{\beta (\varepsilon_{\textbf{q}} -\mu)}- 1} \  ,
\end{equation}
where for ideal {\textit{bosons}}: {$\mu <0$}, since
$\varepsilon_{k} =\hbar^2 \textbf{k}^2/(2m)\geq0$, $\textbf{k} \in \Lambda^*$, by virtue of
(\ref{k-Lambda}).

Owing to (\ref{k-Lambda}) and (\ref{q-dens}) the {grand-canonical} expectation of the
{\textit{total}} density of bosons in $\Lambda$ is
\begin{equation}\label{1-g-c-densV}
\rho_{\Lambda}(\beta,\mu):= \frac{1}{V}\sum_{\textbf{k}\in\Lambda^{\ast}}
\frac{1}{e^{\beta (\varepsilon_{\textbf{k}} -\mu)}- 1} =
\frac{1}{L^3}\sum_{\{n_{j}\, \in \ \mathbb{Z}: \, j=1,2,3\}}
\left\{e^{\beta (\hbar^2 \sum_{j=1}^{3} (2\pi n_{j}/L)^2/2m - \mu)}- 1\right\}^{-1}.
\end{equation}

(a) To study the values of the chemical potential for observables, one first considers the
\textit{equation} for $\mu$ and a given total particle density $\rho$ in a \textit{finite} volume
$V = |\Lambda|$:
\begin{equation}\label{2-g-c-densV}
\rho = \rho_{\Lambda}(\beta,\mu) \, , \quad \mu < 0.
\end{equation}
Seeing that due to the term $\{\textbf{k}=0\}$ in (\ref{1-g-c-densV}) the limit
$\lim_{\mu \rightarrow 0} \rho_{\Lambda} (\beta,\mu)= + \infty$, the solution
$\mu_{\Lambda}(\beta, \rho)$
of equation (\ref{2-g-c-densV}) \textit{always} exists and $\mu_{\Lambda}(\beta, \rho)<0$ for
$\rho \geq 0$.
As a consequence, there is no \textit{macroscopic} (proportional to the volume) occupation
of any single mode $\textbf{k} \in \Lambda^*$, see (\ref{q-dens}) for $\mu = \mu_{\Lambda}(\beta, \rho)$,
and, thus (as it was noticed by Uhlenbeck in \cite{Uhlenbeck}) there is \textit{no trace} of the BEC,
or of any phase transitions.

At that time this conclusion was not anymore surprising \cite{KahnUhlenbeck}, since after the
"100th anniversary of Van der Waals' birthday Congress" (Amsterdam 1937) the idea, that
condensation as a phase transition could mathematically be understood only in the
\textit{thermodynamic limit} $V \rightarrow \infty$, became dominating.

(b) Let $\mu< 0$ and $\Lambda \nearrow \mathbb{R}^{d=3}$. Note that the last term in
(\ref{1-g-c-densV}) is nothing but the integral \textit{Darboux-Riemann} sum, which converges
for $L \rightarrow \infty$ to integral $\mathfrak{I}_{d=3}(\beta,\mu)$:
\begin{eqnarray}\label{3-g-c-densV}
\rho(\beta,\mu) := \lim_{\Lambda}{\rho_\Lambda (\beta,\mu)}
= \frac{1}{(2\pi)^3}\int_{\mathbb{R}^3} d^3 k \left\{e^{\beta(\hbar^2 k^2/2m -\mu)}- 1\right\}^{-1}
=:\mathfrak{I}_{d=3}(\beta,\mu).
\end{eqnarray}
Note that for $d >2$ the integral $\mathfrak{I}_{d}(\beta,\mu)$  is convergent and bounded for
$\mu \leq 0$ with $\sup_{\mu \leq 0} \mathfrak{I}_{d=3}(\beta,\mu) = \rho(\beta,\mu =0) =:
\rho_c(\beta)$. Hence we face up to the fact that in the \textit{thermodynamic limit} equation
(\ref{2-g-c-densV}) gets the form $\rho = \rho(\beta,\mu)$,  $\mu \leq 0$, and has solutions
$\mu(\beta, \rho)$ \textit{only} for densities less than the critical density:
$\rho \leq \rho_c(\beta)$.
This observation, which looks as a \textit{defect} of the model, which called the 
\textit{ideal} Bose-gas
\cite{Kvasnikov}, Ch.III, has been translated by Uhlenbeck in \cite{Uhlenbeck} as a \textit{no-go}
{assertion} about {possibility} of the BEC phase transition even after the {thermodynamic limit}.

(c) An elegant way to resolve the paradox in (b) and to obtain BEC was suggested by
F. London \cite{London} (1938). Formally his arguments could be presented as follows.
If we search for solutions of equation (\ref{2-g-c-densV}) for $\mu < 0$ in the limit
$V \rightarrow \infty$, then the mathematical problem is to describe a \textit{family} of solutions
of (\ref{2-g-c-densV}) for the sequence of \textit{unbounded} for $\mu \in (-\infty, 0)$ functions
$\{\mu \mapsto \rho_\Lambda (\beta,\mu)\}_{\Lambda}$, which \textit{non-uniformly} in
$\mu$ converges to a \textit{bounded} in $(-\infty, 0)$ function $\rho(\beta,\mu)$.

Since a \textit{singularity} preventing the uniform convergence is due to the $\{\textbf{k}= 0\}$-term
in the right-hand side of (\ref{1-g-c-densV}), we re-write equation (\ref{2-g-c-densV})
as follows:
\begin{equation}\label{4-g-c-densV}
\rho = \frac{1}{L^3}
\left\{ e^{- \beta \mu} - 1 \right\}^{-1} +
\frac{1}{L^3}\sum_{\{n_{j}\, \in \ \mathbb{Z}: \, j=1,2,3\}\backslash \{\textbf{0}\}}
\left\{e^{\beta (\hbar^2 \sum_{j=1}^{3} (2\pi n_{j}/L)^2/2m - \mu)}- 1\right\}^{-1} \, .
\end{equation}
It is important to remark that for $L \rightarrow \infty$ the \textit{Darboux-Riemann} sum in the
right-hand side of (\ref{4-g-c-densV}) converges to integral
$\mathfrak{I}_{d > 2}(\beta,\mu) \leq \rho_c(\beta)$ (\ref{3-g-c-densV}) \textit{uniformly}  in
$\mu \leq 0$.\\
(I) Case $\rho < \rho_c(\beta)$. \\
Then the \textit{Darboux-Riemann} sum in the
right-hand side of (\ref{4-g-c-densV}) is also less than $\rho_c(\beta)$ where solutions
of equation (\ref{4-g-c-densV}) are $\mu_\Lambda (\beta,\rho) < 0$.
Seeing that for $L \rightarrow \infty$ this sum converges uniformly in $\mu < 0$ to integral
(\ref{3-g-c-densV}), which is less than $\rho_c(\beta)$, we deduce that the limit
$\lim_{L \rightarrow \infty} \mu_{\Lambda} (\beta,\rho) = \mu(\beta,\rho) < 0$. Therefore,
this strictly negative limit is solution for (\ref{3-g-c-densV}): $\rho = \rho(\beta, \mu(\beta,\rho)$,
whereas for another term in (\ref{4-g-c-densV}) we obtain
$\lim_{L \rightarrow \infty} L^{-3}(\exp(-\beta \mu_{\Lambda} (\beta,\rho)) -1)^{-1} =0$.\\
(II) Case $\rho > \rho_c(\beta)$. \\
For these values of the total density $\rho$ the volume-dependent \textit{family} of solutions
$\{\mu_\Lambda (\beta,\rho)\}_{\Lambda}$  of equation (\ref{4-g-c-densV}) satisfies the identity
\begin{equation}\label{5-g-c-densV}
\rho - \frac{1}{L^3}\sum_{\{n_{j}\, \in \ \mathbb{Z}: \, j=1,2,3\}\backslash \{\textbf{0}\}}
\left\{e^{\beta (\hbar^2 \sum_{j=1}^{3} (2\pi n_{j}/L)^2/2m - \mu_\Lambda (\beta,\rho))}- 1\right\}^{-1}
 = \frac{1}{L^3} \left\{ e^{- \beta \mu_\Lambda (\beta,\rho)} - 1 \right\}^{-1}
 \, .
\end{equation}
Owing to the facts: (a) that the \textit{Darboux-Riemann} sum in the right-hand side of (\ref{4-g-c-densV})
for $\mu \leq 0$ is less than $\rho_c(\beta)$ and (b) that for $L \rightarrow \infty$ it converges to
integral $\mathfrak{I}_{d > 2}(\beta,\mu) \leq \rho_c(\beta)$ (\ref{3-g-c-densV}) \textit{uniformly}  in
$\mu \leq 0$, we deduce the limits:
\begin{equation}\label{6-g-c-densV}
\lim_{L \rightarrow \infty}
\frac{1}{L^3}\sum_{\{n_{j}\, \in \ \mathbb{Z}: \, j=1,2,3\}\backslash \{\textbf{0}\}}
\left\{e^{\beta (\hbar^2 \sum_{j=1}^{3} (2\pi n_{j}/L)^2/2m - \mu_\Lambda (\beta,\rho))}- 1\right\}^{-1}
 = \mathfrak{I}_{d > 2}(\beta,\lim_{L \rightarrow \infty}\mu_\Lambda (\beta,\rho)))
\leq \rho_c(\beta)  \, .
\end{equation}
Then by condition $\rho > \rho_c(\beta)$ and equations (\ref{5-g-c-densV}) and (\ref{6-g-c-densV})
one gets that solutions $\{\mu_\Lambda (\beta,\rho)\}_{\Lambda}$, for $L \rightarrow \infty$,
\textit{must} converge to \textit{zero} and that
\begin{equation}\label{7-g-c-densV}
\rho - \rho_c(\beta) =
\lim_{L \rightarrow \infty}\frac{1}{L^3}
\left\{ e^{- \beta \mu_\Lambda (\beta,\rho)} - 1 \right\}^{-1} > 0
 \, .
\end{equation}
- As a consequence, (\ref{7-g-c-densV}) yields that for $\rho > \rho_c(\beta)$ there is a
\textit{macroscopic} occupation of the mode $\{\textbf{k}= 0\}$
\begin{equation}\label{8-g-c-densV}
\rho_0 (\beta) := \lim_{V\rightarrow\infty}\frac{\left\langle N_{k=0}\right\rangle_{T_{\Lambda }}
(\beta,\mu_\Lambda (\beta,\rho))} {V} = \rho - \rho_c(\beta) > 0 \, ,
\end{equation}
that is, the {Bose-Einstein condensation} in the \textit{zero-mode}. \\
- Moreover, (\ref{7-g-c-densV}) allows to estimate the \textit{rate} of convergence to {zero}
of solutions $\{\mu_\Lambda (\beta,\rho)\}_{\Lambda}$:
\begin{equation}\label{9-g-c-densV}
\mu_{\Lambda}(\beta,\rho > \rho _{c}(\beta))=  {- \, \frac{1}{L^3}} \ \frac{1}
{\beta(\rho -\rho _{c}(\beta))} + {o}({1}/{V}) \, , \quad L \rightarrow \infty \, .
\end{equation}
Taking into account (\ref{9-g-c-densV}) one can check the occupation density of \textit{non-zero} modes.
By virtue of $\{\textbf{k} = (2\pi /L) \{n_1, n_2, n_3\} \neq \textbf{0}\}$ and by asymptotics
(\ref{9-g-c-densV}) we obtain  \, .
\begin{eqnarray}\label{10-g-c-densV}
&&\lim_{V\rightarrow\infty}\frac{\left\langle N_{\textbf{k}}\right\rangle_{T_{\Lambda }}
(\beta,\mu_\Lambda (\beta,\rho))} {V} = \lim_{L \rightarrow \infty}\frac{1}{L^3}
\{\exp(\beta (\hbar^2 \sum_{j=1}^{3} (2\pi n_{j}/L)^2/2m - \mu_\Lambda (\beta,\rho)))-
1\}^{-1} = \\
&&\lim_{L \rightarrow \infty}\frac{1}{L^3}
\{\exp(\beta (\hbar^2 \sum_{j=1}^{3} (2\pi n_{j}/L)^2/2m
+ \, {L^{-3}} \, ({\beta(\rho -\rho _{c}(\beta))})^{-1} - {o}({1}/{L^{3}}))-
1\}^{-1} = \nonumber \\
&&\lim_{L \rightarrow \infty}\frac{1}{L^3} \{\exp(\beta \hbar^2 \sum_{j=1}^{3} (2\pi n_{j}/L)^2/2m)-
1\}^{-1} = 0
 \ .
\end{eqnarray}
Hence, for \textit{any} density of bosons $\rho$ there is no BEC in \textit{any} of \textit{non-zero}
modes $\{\textbf{k} \neq \textbf{0}\}$.
\\
(III) Case $\rho = \rho_c(\beta)$. \\
The analysis of this case is more delicate \cite{ZagBru01}, Ch.2. Instead of (\ref{9-g-c-densV})
one gets asymptotics $\mu_{\Lambda}(\beta,\rho _{c}(\beta)) \simeq \mathcal{O}(1/V^{\alpha})$,
for some $\alpha \in (2/3,1)$.

Summarising, the ideal Bose-gas in \textit{cubic} box $\Lambda =L\times L\times L \subset \mathbb{R}^{d=3}$,
for a given temperature and for particle densities larger than critical, manifests the one-mode
Bose-Einstein condensation (\ref{8-g-c-densV}) in the ground zero-mode state for
one-particle kinetic-energy operator with periodic self-adjoint extension.
There are straightforward generalisations of this assertion to dimensions $d > 2$, as well as to any
bounded convex $\Lambda \subset \mathbb{R}^d$ with smooth boundary $\partial \Lambda$ and to different
self-adjoint extensions with \textit{attractive/non-attractive} boundary conditions, see
\cite{ZiffUhlenbeckKac}, \cite{Verbeure}.

\vspace{0.2cm}

After 1938 new developments in the theory of BEC in continuous \textit{translation-invariant}
systems emerged. The first one in 1982-1986, due to {van den Berg-Lewis-Pul\`{e}} (a \textit{generalised}
BEC of Types I, II, III), the second one a bit later, when van den Berg-Lewis-Pul\`{e} (1988)
\cite{BergLewisPule1} discovered in the \textit{Huang-Yang-Luttinger} (HYL) model \cite{HuangYangLuttinger}
for \textit{interacting} bosons a \textit{non-conventional} BEC, which exists \textit{because of}
interaction. Later, such unusual {non-conventional} BEC was also found in the
\textit{Bogoliubov weakly-imperfect Bose-gas } (1998)\cite{BruZagrebnov1}.

Nowadays the \textit{one-mode} BEC (\ref{8-g-c-densV}) is known as \textit{conventional}
{Bose-Einstein condensation of the {Type} {I}}, see \cite{BergLewisPule2}.
We shall return to the \textit{generalised} BEC and to the \textit{non-conventional} BEC
in the next subsections.

\subsubsection{Generalised (Bose-Einstein) condensation
\textit{\`{a} la} van den Berg-Lewis-Pul\`{e}} \label{subsubsection 1.2.2}

The simplest way to understand different types of \textit{generalised} BEC is to consider the example
motivated by \textit{Hendrik Casimir} \cite{BergLewis1}(1968). Let us instead of the \textit{cubic box}
$\Lambda =L\times L\times L \subset \mathbb{R}^3$,
$\left|\Lambda \right| =V$, take a \textit{prism} $\Lambda =L_{1}\times L_{2}\times L_{3}$
of the \textit{same} volume with the sides of length $L_{j}=V^{\alpha _{j}}$, $j=1,2,3,$, such that
$\alpha _{1}\geq \alpha_{2}\geq \alpha _{3}>0$
and $\alpha _{1}+\alpha _{2}+\alpha _{3}=1$. Here we {ignore} a conflict between \textit{linear}
and \textit{volume} dimensionalities as irrelevant for further calculations.

\vspace{0.3cm}
\noindent \textbf{Generalised BEC Type} $\mathbf{II}$ \\
Let $\Lambda$ be the \textit{anisotropic Casimir} box, that is, $\alpha _{1}=1/2.$. Since for
$\{\textbf{k} = (2\pi) \{n_1/V^{1/2}, 0, 0\}\}_{n_1\in \mathbb{Z}}$ (\ref{k-Lambda}) one gets
$\varepsilon_{k}= (2\pi\hbar \ n_1)^{2}/{2m V}$  we re-write equation (\ref{4-g-c-densV}) as follows:
\begin{eqnarray}
\rho &=&\frac{1}{V}\frac{1}{e^{-\beta \mu }-1}+\frac{1}{V}{\sum_{k\in
\left\{ \Lambda ^{\ast }:n_{1}\neq 0,n_{2}=n_{3}=0\right\} }}\frac{1}{e^{\beta
\left( \varepsilon _{k}-\mu \right) }-1}
\nonumber \\
&&+\frac{1}{V}{\sum_{k\in \left\{ \Lambda ^{\ast}:n_{j}\neq 0,j=2\text{
or }3\right\} } }\frac{1}{e^{\beta \left( \varepsilon _{k}-\mu \right)}-1} \ . \label{1-gBEC}
\end{eqnarray}

Note that for $\mu < 0$ the \textit{second} term in the right-hand side of (\ref{1-gBEC}) is the
\textit{Darboux-Riemann} sum for \textit{one-dimensional integral} divided by $V^{\alpha _{2}+\alpha _{3}}$
and, thus, tends to \textit{zero} for $V \rightarrow \infty$. The \textit{third} term in the right-hand
side of (\ref{1-gBEC}) is the \textit{Darboux-Riemann}
sum for \textit{three-dimensional integral} (\ref{3-g-c-densV}), which, as above, converges, for
$V \rightarrow \infty$, to $\mathfrak{I}_{3}(\beta,\mu) \leq \rho_c(\beta)$  \textit{uniformly} in
$\mu \leq 0$. If $\{\mu_\Lambda (\beta,\rho)\}_{\Lambda}$ are solutions of equation (\ref{1-gBEC}),
then it yields the limits:
\begin{equation}\label{2-gBEC}
{\lim_{V \rightarrow \infty}} \left\{\frac{1}{V}\frac{1}{e^{-\beta \mu_{\Lambda }\left( \beta ,
\rho \right) }-1}
+ \frac{1}{V}{\sum_{k\in
\left\{ \Lambda ^{\ast }:n_{1}\neq 0,n_{2}=n_{3}=0\right\}}}
\frac{1}{e^{\beta \left( \varepsilon _{k}-\mu _{\Lambda }\left(\beta ,\rho
\right) \right) }-1} \right\}=\rho -
{\lim_{V \rightarrow \infty}}\mathfrak{I}_{3}(\beta,\mu_{\Lambda }\left( \beta ,\rho \right)) .
\end{equation}
Note that if $\mu_\Lambda (\beta,\rho) < 0$, then the limit in the left-hand side of (\ref{2-gBEC}) is
\textit{zero}. Therefore, to satisfy (\ref{2-gBEC}) for $\rho > \rho _{c}(\beta)$ the solutions
$\{\mu_\Lambda (\beta,\rho)\}_{\Lambda}$ of equation (\ref{1-gBEC}) \textit{must} converge to zero:
$\lim_{V \rightarrow \infty} \mu_{\Lambda }\left( \beta ,\rho \right) =0$. This may ensure that the
limit in the left-hand side yields a positive difference:
$\rho - {\lim_{V \rightarrow \infty}}\mathfrak{I}_{3}(\beta,\mu_{\Lambda }\left( \beta ,\rho \right)) =
\rho - \rho_c(\beta) > 0$. On account of $\alpha _{1}=1/2$ one obtains in the left-hand side of
(\ref{2-gBEC}) that $\varepsilon_{k}= (2\pi\hbar \ n_1)^{2}/{2m V}$
for $\{\textbf{k} = (2\pi) \{n_1/V^{1/2}, 0, 0\}\}_{n_1\in \mathbb{Z}}$. Hence,
a non-zero limit in the left-hand side of (\ref{2-gBEC}) implies for solutions
$\mu _{\Lambda }\left( \beta ,\rho >\rho _{c}\left( \beta \right) \right)$ the asymptotic behaviour:
\begin{equation}\label{3-gBEC}
\mu _{\Lambda }\left( \beta ,\rho >\rho _{c}\left( \beta \right)
\right) = - \frac{A}{V}+o\left( \frac{1}{V}\right) ,\text{ }A>0 \, .
\end{equation}
Then asymptotics (\ref{3-gBEC}) suggests for the left-hand side of (\ref{2-gBEC}) the non-zero limit:
\begin{equation}\label{4-gBEC}
{\lim_{V \rightarrow \infty}} \ \frac{1}{V}  {\sum_{k\in \left\{ \Lambda ^{\ast}:\left( n_{1},0,0\right)
\right\}}}
\frac{1}{e^{\beta \left(\varepsilon_{k}-\mu _{\Lambda }\left( \beta ,\rho \right) \right) }-1}
=\frac{1}{\beta }{\sum_{n_{1}=0,\pm 1,\pm 2,...}}
\left(\frac{\left( 2\pi \hslash \right)^{2}}{2m}n_{1}^{2}+A\right) ^{-1}.
\end{equation}

As a consequence, the limits (\ref{2-gBEC}), (\ref{4-gBEC}) provide a \textit{generalised} BEC
$\rho_0(\beta)$  with \textit{macroscopic} occupation of a \textit{countable} number of modes:
\begin{equation}\label{5-gBEC}
\rho_0(\beta) = \frac{1}{\beta }{\sum_{n_{1}=0,\pm 1,\pm 2,...}}
\left(\frac{\left(2\pi \hslash \right)^{2}}{2m}n_{1}^{2}+
A\right)^{-1}=\rho -\rho_{c}\left(\beta \right) \, .
\end{equation}
Here parameter $A=A(\beta ,\rho)$ (\ref{3-gBEC}) is a \textit{unique} root of equation (\ref{5-gBEC}).
Moreover, seeing that $\alpha_2 + \alpha_3 =1/2$, the limit (\ref{4-gBEC}) with $\varepsilon_{k}$ for
other modes: $\left\{ \Lambda^{\ast }:n_{j}\neq 0, j=2, 3 \right\}$ is \textit{nulle} because
$\varepsilon_{k}\sim \mathcal{O}(V^{- 2 \alpha_2}) \ {\rm{or}} \ \mathcal{O}(V^{- 2 \alpha_3})$,
whereas $\mu _{\Lambda }\left( \beta ,\rho >\rho _{c}\left( \beta \right)\right) \sim \mathcal{O}(V^{- 1})$
(\ref{3-gBEC}).
This shows that for $\rho >\rho _{c}\left( \beta \right)$ only modes
$\left\{ \Lambda ^{\ast }:\left(n_{1},0,0 \right) \right\}$ are \textit{macroscopically} occupied:
\begin{eqnarray}\label{6-gBEC}
&&{\lim_{V \rightarrow \infty}} \ \frac{1}{V} \ \left\langle N_{k}\right\rangle_{T_{\Lambda }}
\left(\beta,\mu_{\Lambda }
\left( \beta ,\rho \right)\right) = \\
&&\left\{
\begin{array}{l}
\beta^{-1}\left({\left(2\pi \hslash n_{1}\right)^{2}}/{2m} +
A(\beta ,\rho) \right) ^{-1},\text{ for }k\in \left\{ \Lambda ^{\ast }:\left(n_{1},0,0\right)
\right\} \\
0\text{ }\quad \text{\quad \quad \quad \quad \quad \quad \quad \quad \quad} \ , \ \text{ for }
k\in \left\{ \Lambda ^{\ast }:n_{j}\neq 0,j=2 , 3\right\}
\end{array}
\right\} .  \nonumber
\end{eqnarray}

This means that for a \textit{long} anisotropic prism with $\alpha _{1}=1/2$ in the
\textit{thermodynamic limit} ($L \rightarrow \infty$) there exists  \textit{macroscopic }occupation
of an \textit{infinite }number of low-lying modes
$k\in \left\{ \Lambda ^{\ast }:\left( n_{1},0,0\right) \right\} $ including the \textit{zero-mode} $k=0$.
In contrast to the \textit{Type} I BEC, which occupies one, or few,  modes, this is the case of the
{van den Berg-Lewis-Pul\`{e}} \textit{generalised} BEC of \textit{Type} II (1986) \cite{BergLewisPule2}.

\vspace{0.3cm}
\noindent \textbf{Generalised BEC Type} $\mathbf{III}$ \\
Now let $\alpha _{1}>1/2$. That is, we consider a \textit{highly anisotropic} prism still in one
direction $j=1$ \cite{BergLewisPule2}. Then
\begin{equation}
{\inf_{k\in \Lambda^{\ast}\backslash\left\{0\right\}}}
\varepsilon _{k}=\frac{\left( 2\pi \hslash \right) ^{2}}{2m}\frac{1}{%
V^{2\alpha _{1}}}\, , \quad 2\alpha _{1}>1,  \label{Physreport97}
\end{equation}
which corresponds to the mode with $\left( n_{1}=1,n_{2}=0,n_{3}=0\right) $.
Since for any $\mu <0$ the right-hand side of (\ref{1-gBEC}) converges
to the integral $\rho \left( \beta ,\mu \right) $ monotonously
increasing up to $\rho _{c}\left( \beta \right) $ for  $\mu
\rightarrow -0,$ the solution $\mu _{\Lambda }\left( \beta ,\rho
>\rho _{c}\left( \beta \right) \right) $ of (\ref{1-gBEC}) has (for $V\rightarrow \infty$) the asymptotics :
\begin{equation}
\mu _{\Lambda }\left( \beta ,\rho >\rho _{c}\left( \beta \right)
\right) =- \frac{B}{V^{\delta }}+o\left( \frac{1}{V^{\delta }}\right) ,\text{ }%
B>0,\text{ }\delta >0.  \label{Physreport98}
\end{equation}
For calculation $B$ and $\delta ,$ we remark that the
first \textit{two} terms in the right-hand side of (\ref{1-gBEC}) may be represented as:
\begin{eqnarray}
&&\frac{1}{V}{\sum_{k\in \left\{ \Lambda ^{\ast }:\left( n_{1},0,0\right)
\right\}}}\frac{1}{e^{\beta \left( \varepsilon _{k}-\mu _{\Lambda}\left(\beta,\rho \right)\right) }-1}
=\frac{1}{V}{\sum_{k\in\left\{ \Lambda ^{\ast }:\left( n_{1},0,0\right) \right\}}} \ \
{{\sum_{s=1}^{+\infty}}}e^{-s\beta \left( \varepsilon
_{k}-\mu _{\Lambda }\left( \beta ,\rho \right) \right)}=  \nonumber \\
&&\frac{1}{V} \ {\sum_{s=1}^{+\infty}}e^{s\beta \mu_{\Lambda }\left( \beta ,\rho \right) } \ \
{\sum_{n_{1}=0,\pm 1,\pm 2,...}}e^{-s\beta \left( {{\left( 2\pi \hbar  n_{1}\right)^{2}}}/{2m}
{V^{2\alpha _{1}}}\right) } \ .  \label{Physreport99}
\end{eqnarray}
Notice that the \textit{Jacobi identity} gives for the last sum in (\ref{Physreport99}):
\begin{equation}
{\sum_{n_{1}=0,\pm 1,\pm 2,...} }e^{-\pi \lambda n_{1}^{2}}=
\frac{1}{\sqrt{\lambda }}{\sum_{\xi =0,\pm 1,\pm 2,...}}e^{-\left( \pi
\xi ^{2}/\lambda \right) },  \label{Physreport100}
\end{equation}
where $\lambda =s\beta \, {2\pi \hslash ^{2}}V^{-2\alpha _{1}}/{m}.$ Taking
into account (\ref{1-gBEC}) and (\ref{Physreport98})-(\ref{Physreport100}) we find that for
$\lambda \rightarrow 0$ only the term with $\xi =0$ is important for (\ref{Physreport99})
when $V\rightarrow \infty $, and the limit (\ref{2-gBEC}) takes the form:
\begin{equation}
\rho -\rho _{c}\left( \beta \right) ={\lim_{V\rightarrow \infty}
}\left\{ \left( \frac{2\pi \hslash ^{2}}{m}\beta \right) ^{-1/2}\left\{
\frac{V^{\alpha _{1}-1}}{V^{\delta /2}}\cdot V^{\delta }\right\} \frac{1}{%
V^{\delta }}\left\{ {{\sum_{s=1}^{+\infty}}}e^{- \beta
B\left( s/V^{\delta }\right) }\left( \frac{s}{V^{\delta }}\right)
^{-1/2}\right\} \right\} .  \label{Physreport101}
\end{equation}
By inspection the right-hand side of (\ref{Physreport101}) with the \textit{Darboux-Riemann} sum is
\textit{nontrivial} only for
\begin{equation}
\delta =2\left( 1-\alpha _{1}\right) .  \label{Physreport102}
\end{equation}
Then for $\rho -\rho _{c}\left( \beta \right) >0 $ one gets
\begin{equation}
\rho -\rho _{c}\left( \beta \right) =\left( \frac{2\pi \hslash ^{2}}{m%
}\right) ^{-1/2}\beta ^{-1/2}{{\int_{0}^{+\infty}}} d\xi
e^{- \beta B\xi }\xi ^{-1/2},  \label{Physreport103}
\end{equation}
where $B=B\left( \beta,\rho \right) > 0$ is the \textit{unique} root of equation (\ref{Physreport103}),
that is,
\begin{equation}
\rho -\rho _{c}\left( \beta \right) =\left( \frac{m}{2\hslash ^{2}}
\right) ^{1/2}\frac{1}{\beta \sqrt{ B\left( \beta,\rho \right)}}.  \label{Physreport104}
\end{equation}
Thanks to (\ref{Physreport98}), (\ref{Physreport102}) and (\ref
{Physreport104}) we obtain that for $\rho >\rho _{c}\left( \beta \right) $
and $V\rightarrow \infty $ the asymptotics:
\begin{equation}
\varepsilon _{k}-\mu _{\Lambda }\left( \beta ,\rho \right) \simeq
\left\{
\begin{array}{l}
(\hslash ^{2}/{2m})\left( {2\pi n_{1}}/{V^{\alpha _{1}}}\right) ^{2}+
(m/2\hslash ^{2})/(\beta ^{2}\left( \rho -\rho _{c}\left( \beta
\right) \right) ^{2}V^{2\left( 1-\alpha _{1}\right) }), \\ \text{ }k\in \left\{
\Lambda ^{\ast }:\left( n_{1},0,0\right) \right\} \\
(\hslash ^{2}/{2m})\left[ \left( {2\pi n_{2}}/{V^{\alpha _{2}}}\right)^{2}+
\left( {2\pi n_{3}}/{V^{\alpha _{3}}}\right)^{2}\right] + \\
(m/2\hslash ^{2})/(\beta ^{2}\left( \rho -\rho _{c}\left( \beta
\right) \right) ^{2}V^{2\left( 1-\alpha _{1}\right) }),
\text{ }k\in \left\{\Lambda ^{\ast }:n_{j=2\text{ or }3}\neq 0\right\}
\end{array}
\right\} .  \label{Physreport105}
\end{equation}

Since $\alpha _{1}>1/2$ and $\alpha _{1}+\alpha _{2}+\alpha _{3}=1$, the
asymptotics (\ref{Physreport105}) imply
\begin{equation}
{\lim_{V\rightarrow \infty}} \ \frac{1}{V}\left\langle N_{k}\right\rangle
_{T_{\Lambda }}\left( \beta ,\mu _{\Lambda }\left( \beta ,\rho >\rho
_{c}\left( \beta \right) \right) \right) =0,\text{ }k\in \Lambda ^{\ast
},  \label{Physreport106}
\end{equation}
and at the same time
\begin{equation}\label{gBEC1}
{\lim_{V\rightarrow \infty}} \ \frac{1}{V^{2\left( 1-\alpha _{1}\right) }}%
\left\langle N_{k}\right\rangle _{T_{\Lambda }}\left( \beta ,\mu _{\Lambda
}\left( \beta ,\rho >\rho _{c}\left( \beta \right) \right) \right)
=\frac{m/2\hslash ^{2}}{\beta ^{2}\left( \rho -\rho _{c}\left( \beta
\right) \right) ^{2}},\text{ }k\in \left\{ \Lambda ^{\ast }:\left(
n_{1},0,0\right) \right\},
\end{equation}
thought by (\ref{1-gBEC}), (\ref{Physreport99}), (\ref{Physreport101})) we deduce for density of
\textit{generalised} BEC:
\begin{equation}\label{gBEC2}
\rho _{0}\left( \beta \right) =\rho -\rho _{c}\left( \beta \right)
={\lim_{V\rightarrow \infty}} \ \frac{1}{V}{\sum_{k\in \left\{
\Lambda ^{\ast }:\left( n_{1},0,0\right) \right\}}}\left\langle
N_{k}\right\rangle _{T_{\Lambda }}\left( \beta ,\mu _{\Lambda }\left(
\beta ,\rho \right) \right) >0.
\end{equation}
This means that in the case of \textit{extremely} {long} prism with $\alpha_{1}>1/2$ there exists
for $\rho >\rho _{c}\left( \beta \right) $ a \textit{conventional generalised} BEC
with density $\rho _{0}\left( \beta \right) >0$ (\ref{gBEC2}), which is according to
van den Berg-Lewis-Pul\'{e} of the \textit{Type} III  \cite{BergLewisPule2},
because there is \textit{no any} macroscopically occupied level in $\Lambda ^{\ast }$,
see (\ref{Physreport106}) and (\ref{gBEC1}).

These observations give a motivation for the following van den Berg-Lewis-Pul\'{e}'s
{\textit{classification}} of the \textit{generalised} BEC \cite{Berg1,BergLewis1,BergLewisPule2} :

\begin{itemize}
\item  the condensation is called the \textit{Type I} if a \textit{finite} number
of single-particle levels are macroscopically occupied;

\item  it is of \textit{Type II} if an \textit{infinite} number of the levels are
macroscopically occupied;

\item  it is called the \textit{Type III}, or the \textit{\textit{%
non-extensive}} condensation, if \textit{no} of the levels are macroscopically
occupied , whereas one has :
\begin{equation}\label{def-gBEC}
\rho_{0}\left(\beta \right) = {\lim_{\delta \rightarrow 0^{+}}}{\lim_{V\rightarrow \infty}} \
\frac{1}{V}{\sum_{\left\{ k\in \Lambda ^{\ast },0\leq \left\| k\right\|
\leq \delta \right\}}}\left\langle N_{k}\right\rangle =\rho -\rho_{c}\left(\beta \right) .
\end{equation}
This double limit (\ref{def-gBEC}) in the van den Berg-Lewis-Pul\'{e} \textit{definition} of the
condensed fraction of bosons, which includes all \textit{Types} of BEC in modes
$\{ k\in \Lambda ^{\ast }\}$, cf. (\ref{gBEC2}).
\end{itemize}
After the Casimir example \cite{BergLewis1} van den Berg-Lewis-Pul\`{e}
\cite{Berg2}, \cite{BergLewisPule2} discovered that BEC in \textit{exponentially}
anisotropic boxes may produce a new phenomena: the \textit{second critical} density
$\rho_m(\beta) > \rho_c(\beta)$ and a quite unusual transition between generalised condensations of the
\textit{Type} I and the \textit{Type} III.
This observation was then confirmed also for other types of \textit{exponentially} anisotropic particle
confinements in \cite{BZ2010}.

\subsubsection{Non-conventional (Bose-Einstein) condensation} \label{subsubsection 1.2.3}

According to subsections \ref{subsubsection 1.2.1} and \ref{subsubsection 1.2.2} the BEC of \textit{Type} I
(\ref{8-g-c-densV}), as well as, the BEC of \textit{Type} II (\ref{5-gBEC}) and of \textit{Type} III
(\ref{gBEC2}) appear in the ideal Bose-gas for $\rho > \rho_c (\beta)$ due to the
\textit{saturation} mechanism owing to the \textit{finite} value of the \textit{critical} density
$\rho_c (\beta)< \infty$.\\
(We remind that this terminology is related to phenomenon when saturated water vapor ("Bose-gas")
will condense to form liquid water called \textit{dew} ("condensate"). It happens when density of vapor
is growing and to become \textit{saturated} at the critical vapor density, known as the 
\textit{dew-point}.)

\vspace{0.2cm}
Although the BEC, or \textit{generalised} BEC (\ref{def-gBEC}) in the ideal Bose-gas, were
studied in a great details, analysis of condensate in the \textit{interacting} Bose-gas is a more 
delicate
problem. Recall that \textit{effective statistical} attraction between bosons (see
subsection \ref{subsubsection 1.1.2}), which is behind of the BEC for the ideal Bose-gas,
makes this system unstable with respect to any \textit{direct} \textit{attractive} interaction between
bosons. So, efforts around the question: "Why do interacting bosons condense?", were essentially 
concentrated
around \textit{repulsive} interactions between bosons. The studying of stability of the
\textit{conventional} BEC (or gBEC) in the non-ideal Bose-gas with a rapidly decaying direct two-body
\textit{repulsive} interaction is still in progress, see, e.g.,  \cite{SU09,BU10}.
On the other hand, if one counterbalances direct (\textit{and statistical}) attractive interaction by a
repulsion stabilising the boson system, this attraction may be the origin of a {new}
(\textit{non-saturating}) mechanism of BEC called the \textit{non-conventional} (dynamical) condensation
\cite{ZagPPN2021}. Implicitly the \textit{non-conventional} condensation was considered for the first time
in \cite{BergLewisPule1} on the basis of their rigorous study of condensation in the Huang-Yang-Luttinger
(HYL) model \cite{HuangYangLuttinger}.

We note that it was D.J. Thouless \cite{Thouless}, who presented an instructive "back-of-the-envelope"
calculations, which argue that a new kind of Bose condensation may occur in the HYL model of the
hard-sphere Bose-gas with a \textit{jump} of the condensation density (as a function of the chemical
potential) at the critical point. In \cite{BergDorlasLewisPule1} it was called the "Thouless effect".
Ten years later, the \textit{non-conventional} condensation with a \textit{jump} on the critical
line (for \textit{condensed-noncondensed} phases) was discovered also in the Bogoliubov Weakly Imperfect
Bose Gas (WIBG), see \cite{BruZagrebnov1} and reviews \cite{ZagBru01}, \cite{ZagPPN2021}.

Difference between \textit{conventional} and \textit{non-conventional}
condensations reflects the difference in the mechanism of their formation \cite{BergLewisPule1},
\cite{BergDorlasLewisPule1}:\\
- The \textit{conventional} condensation is a consequence of the balance
between \textit{entropy} and \textit{kinetic energy} via mechanism of saturation. \\
- The \textit{non-conventional} condensation results from the balance between \textit{entropy}
and \textit{interaction energy}, that is, via interaction-induced mechanism.\\
This difference has an important consequence: a non-conventional condensation occurs only \textit{due}
to particle interactions.

The last remark motivates also another name for non-conventional condensation:
the \textit{dynamical} condensation \cite{ BZ98QPL, ZagPPN2021}.
As a consequence, the dynamical condensation may occur in \textit{low-dimensional} ($d\geq 1$)
boson systems, when there is \textit{no} condensation \textit{without} interaction,
as well as, it may exhibit the \textit{first-order} phase transition with a \textit{jump} in the density
of condensate at the critical point (or line). The both HYL and WIBG models manifest these properties, 
which
for the HYL model has been predicted in \cite{Thouless} and then proved in
\cite{BergLewisPule1, BergDorlasLewisPule1}, see \cite{DorlasLewisPule} for further development
and more results in this direction.

For WIBG model the proof of the jump in the density of the \textit{zero-mode} condensate at the critical
line between two phases (condensed-noncondensed) was demonstrated in \cite{BZ98JP} for dimensions $d\geq1$.
Besides the fact of condensation at \textit{low} dimensions, the interaction-induced condensate in the
WIBG model emerges in \textit{two} stages. First, it appears
as \textit{non-conventional} zero-mode condensate with a \textit{jump} ("Thouless effect"), and then
second, as a (continuously growing) \textit{conventional} {generalised} Bose-Einstein condensate
of \textit{Type} I (\textit{out} of the zero-mode !), due to the particles \textit{saturation} mechanism,
see \cite{BZ00} Sections 5.2 and 5.3. Note that for calculation the mentioned above condensate of
\textit{Type} I out of the zero-mode, one must use the van den Berg-Lewis-Pul\'{e} definition
(\ref{def-gBEC}), which is the way to take into account the modes involved into condensation, cf.
\cite{BZ00} Corollary 5.17, and \cite{ZagBru01} Section 5.

\vspace{0.2cm}
In conclusion it is instructive to examine the HYL model in more detail. First we note that it has
Hamiltonian, which is diagonal in the occupation number operators, cf. the kinetic-energy operator
(\ref{Eigenfunc-Spec-N}). Then, it follows that owing to the correspondence established in
subsection \ref{subsubsection 1.2.1}, it is possible to regard the occupation numbers
$\{N_\textbf{k}\}_{\textbf{k} \in \Lambda^*}$ as \textit{random} variables (with values in the natural
numbers and zero) rather than as operators. For example, the variable corresponding to a total number
of bosons in the box $\Lambda$ is $N_{\Lambda} := \sum_{\textbf{k} \in \Lambda^*} N_\textbf{k}$. Hence,
similarly to the kinetic-energy operator (\ref{Eigenfunc-Spec-N}) the Hamiltonian of the HYL model
for $a > 0$ gets the form (cf. \cite{BergLewisPule1}, (1.2) and (1.3))
\begin{eqnarray}\label{HYL}
&&H_{\Lambda }^{HYL}:= H_{\Lambda }^{MF} +
\frac{a}{2V}\{N_{\Lambda }^{2} - {\sum }_{{k\in \Lambda ^{*}}}N_{k}^{2}\} \, , \\
&&H_{\Lambda }^{MF} := {\sum }_{{k\in \Lambda ^{*}}}\varepsilon _{k}\, N_{k}+
\frac{a}{2 V}N_{\Lambda }^{2}\, ,\label{MF}
\end{eqnarray}
where $H_{\Lambda }^{MF}$ is the Hamiltonian of the model with the \textit{mean-field} (MF) interaction.

We recall that repulsive MF-interaction in (\ref{MF}) "improves" the properties of the ideal Bose-gas
in such a way that for chemical potential in this model it is allowed: $\mu \in (-\infty, +\infty)$.
However, MF-model keeps due to the saturation mechanism the \textit{zero-mode} BEC intact with the
\textit{same} as for the IBG critical density $\rho_c(\beta) < \infty$ for $d > 2$
with the total amount of condensate density:
$\rho^{MF}_{0}(\beta, \mu) = \rho(\beta, \mu) - \rho_c(\beta)$,
for $\mu > \mu_{c}^{MF}(\beta)$, see, e.g., \cite{BergLewisSmedt1, LVZ-JSP03}. Here the particle
mean-density is $\rho(\beta, \mu) = \lim_{V \rightarrow\infty}
\left\langle N_{\Lambda}\right\rangle_{H_{\Lambda}^{MF}}\left(\beta,\mu \right)/V$. For the MF-model,
there is \textit{no jump} of the condensate density at the BEC phase transition point 
$\mu_{c}^{MF}(\beta)$,
that is, one gets for the limit:
$\lim_{\mu \rightarrow \mu_{c}^{MF}(\beta) +0} \rho^{MF}_{0}(\beta, \mu) =0 $.

Difference in behaviour of the condensate in models (\ref{HYL}) and (\ref{MF}) reflects a
difference in the origin (mechanism) of the BEC phase transition. In the mean-field model
(\ref{MF}), similarly to ideal Bose-gas, the condensation is a consequence of the balance between entropy
and kinetic energy, which is the first term in the right-hand side of (\ref{MF}). The indicated term is
minimal when all bosons occupy the \textit{zero mode}: $\textbf{k}=0$, and this choice does
\textit{not} affect the interaction term.

On the other hand, the Huang-Yang-Luttinger model (\ref{HYL}) is the MF-interacting Bose-gas (\ref{MF})
(that manifests a \textit{conventional} zero-mode BEC for $d>2$) perturbed by the HYL-interaction term,
the last item in (\ref{HYL}). Then the effect, which favouring the particle accumulation in zero-mode
by the kinetic energy term, is now \textit{enhanced}, for \textit{any} $d\geq 1$,
by this last interaction-energy term since it has the smallest value when \textit{all} bosons occupy the
\textit{same} energy-mode. For that reason, the condensation in the HYL-model is \textit{non-conventional}.
Indeed, it is a result of the balance between entropy and the interaction energy, which produces
(non-conventional) zero-mode BEC even for $d\geq 1$.

This difference has a further consequence. In the \textit{mean-field} model, the \textit{conventional}
condensation occurs if and only if it occurs in the corresponding \textit{ideal} Bose-gas: the mean-field
critical density $\rho_c(\beta)$ is finite only for $d > 2$.
While on the contrary, in the HYL-model due to the particle interaction there is always the
\textit{zero-mode} BEC for sufficiently large density $\rho$ and any $a > 0$. It occurs for any $d \geq 1$
even when for the non-perturbed MF-interacting
boson gas the critical density $\rho_c(\beta)$ for $d \leq 2$ is \textit{infinite} and then it does not
manifest BEC.

\newpage

\vspace{-1.0cm}

\begin{flushright}
\parbox{7.5cm}{``La seule vraie connaissance
est la connaissance des faits.''\\
\textbf{Georges-Louis Buffon,}\textit{\ Histoire naturelle.}
}

\end{flushright}

 \vspace{0.5cm}

\section{JINR-Dubna 1975: Experimental Observation of Bose-Einstein Condensation in the
Superfluid $^{4}$He
 \label{section 2}}


\subsection{Prehistory}\label{subsection 2.1}

The idea to scrutinise and to start experimental study of the Bose-Einstein condensation (BEC) in
superfluid $^{4}\rm{He}$ (He II) was born in Laboratory of Theoretical Physics (JINR) is 1973.
It was during a routine annual meeting of the "Condensed Matter Research Group" chaired by
Professor V.G. Soloviev. Since our colleague V.B. Priezzhev has just defended his PhD Thesis
"Collective excitations in quasi-cristal models of liquids" (Dubna 1973), where essential results
concerned a quasi-cristal model of the liquid Helium ($^{4}\rm{He})$, \cite{Priezzhev-I,Priezzhev-II},
Professor Soloviev asked about applications of these results and suggested
to contact our colleagues from the Laboratory of Neutron Physics. There they possessed a powerful
pulsed reactor INR-30 for \textit{neutronographie} (neutron scattering analysis) of liquid $^{4}\rm{He}$.
This equipment could open an eventual possibility to confirm a long standing hypothesis about existence
of BEC in {He II}, which was a main hypothesis of the \textit{Bogoliubov theory} of superfluidity
\cite{Bogoliubov1, Bogoliubov1bis, Bogoliubov1bisbis}.

So, armed with Priezzhev's thesis and my student's papers \cite{Zag71a, Zag71b} about liquid
$^{4}\rm{He}$ we contacted the Laboratory of Neutron Physics and one of the leading expert
Zh.A. Kozlov, who had been already involved into neutron scattering experiments on liquid
$^{4}\rm{He}$, cf. \cite{Dubnagroup0} and \cite{Kozlov}. The collaboration started by preparation of
theoretical basis for the future experiments.

At that time, one of the interesting for our project theoretical observation was published by
P.C. Hohenberg and P.M. Platzman \cite{HohenbergPlatzman}. They supposed that the \textit{high-energy}
neutrons (with a very short \textit{de Broglie} wave-length) scattered off of single helium atoms may
provide information about the \textit{zero-momentum} BEC in the \textit{superfluid} $^{4}\rm{He}$ (He II).
After \textit{deep-inelastic} scattering from individual "condensed" atoms, the energy transferred
from the neutron would be \textit{almost} equal to the recoil energy \textit{broadened} by (negligible
with respect to the large recoil energy) \textit{final-state} interactions. That is, after being struck
by the high-energy neutrons, the single helium atoms recoil as if they were \textit{free}. While
for atoms \textit{out} of BEC the high-energy transfer would be the recoil energy
\textit{broadened} by the \textit{Doppler shifts} because of non-zero momenta of "non-condensed" atoms.
The (optimistic) estimates given in \cite{HohenbergPlatzman} show that the {Doppler broadening} will be
several times larger that the broadening expected because of the {final-state} interactions. Then neutron
scattering cross sections are anticipated to show \textit{two} components: a \textit{narrow} one for
scattering from the BEC and a \textit{wider} one for scattering from  "non-condensed" atoms of 
$^{4}\rm{He}$.

Before 1974 at least \textit{three} significant experiments, \cite{CowleyWoods}, \cite{Harling} and
\cite{MookSW}, have been carried out to check the neutron scattering suggestion formulated in
\cite{HohenbergPlatzman}.
The results of these papers were severely censured by H.W. Jackson \cite{Jackson} in a long accurate paper
scrutinised these experiments. In spite of in \cite{CowleyWoods}, \cite{Harling}, \cite{MookSW} the
experimental data have been interpreted as giving evidence for a condensate fraction ranging from $2,7\%$
to $17\%$ the conclusion in \cite{Jackson} was that the data of these three experiments appear to be
consistent with the \textit{absence} of the condensate, or no more than a \textit{fraction} of  $1\%$.
This pessimistic conclusion was in a certain conflict with some theoretical estimates of the condensate.

As we have seen in subsections \ref{subsubsection 1.2.2} and \ref{subsubsection 1.2.3}
the notion of the \textit{condensate} of bosons has to be reexamined when we consider the interacting
systems. Therefore, it is appropriate to clarify this notion, since the Bogoliubov theory
\cite{Bogoliubov1, Bogoliubov1bis, Bogoliubov1bisbis}
\textit{insists} on both items: the \textit{condensate} of bosons and the necessity of 
\textit{interaction}
to develop a \textit{microscopic} explanation of superfluidity of the He II. Yet, the \textit{same} 
value of
interaction may completely destroy condensate, which intuitively presented as a fraction of particles,
that does not move, "frozen in the momentum space" with $\textbf{k}=0$.
This picture poses no problem for the perfect Bose-gas, where the particle-number operator 
$N_{\textbf{k}}$
in a given mode, $\textbf{k}\in \Lambda^*$, is the integral of motion. Since single-particle
quantum states are not proper for interacting system, {L. Onsager} and {R. Penrose} (1956)
\cite{Penrose,PenroseOnsager}, worked out a definition of \textit{condensate} in this case. Their proposal
was to identify condensation with an \textit{Off-Diagonal Long-Range Order} (ODLRO), related to
asymptotics of one-particle \textit{density matrix}. This criterion shows that expectation
(\textit{mean-value}) of the occupation number for  \textit{zero-mode} $\textbf{k}=0$ can still be
used as a characterization of the Bose-Einstein condensation. Moreover, they have given about
$8\%$ as estimate of the fraction of particles density in the liquid $^{4}$He having $\textbf{k}=0$ (BEC)
at $T = 0 \,K$.

This was a reason to \textit{revise} in \cite{Dubnagroup01}, \cite{Dubnagroup1} and \cite{Dubnagroup2}
(1973-78) the conclusion of \cite{Jackson} by experiment along the Hohenberg-Platzman suggestions
\cite{HohenbergPlatzman} using a new equipment available at that time in the Laboratory of Neutron Physics
JINR-Dubna. Namely, it was the IBR-30 \textit{pulsed reactor} ensuring the \textit{deep-inelastic} neutron
scattering experiments on the liquid $^{4}$He together with the DIN-1M spectrometer, the time-spectrum
analyser channel was $8 \, \mu$sec.

\subsection{Deep-inelastic neutron scattering and the Bose-Einstein condensate}\label{subsection 2.2}

The neutron-$^{4}\rm{He}$-atoms inelastic \textit{double-differential} scattering cross-section
for $N$ atoms
is given in the first \textit{Born approximation} by formula involved the L. van Hove
\textit{dynamic structure} factor $S(k,\omega)$ \cite{vanHove}:
\begin{equation}\label{1-HP}
\frac{d^2 \sigma}{d\Omega dE_f} = N \, \frac{m_n^2}{(2\pi)^3 \hbar^5} \, \frac{k_i}{k_f} \
|\widehat{U}(k)|^2 \, {S(k,\omega)}
\, .
\end{equation}
One can find more details and explanations in the recent book \cite{AksBal2023} (\S4.1, \S4.2).
Here for neutron with mass $m_n$, initial momentum $\hbar \, \textbf{k}_i$ and energy
$E_i = (\hbar k_i)^2/2 m_n \, $, we denote by $\hbar\textbf{k}= \hbar \textbf{k}_i - \hbar \textbf{k}_f$
and $\hbar \, \omega = E_i - E_f$ the \textit{transferred} momentum and energy corresponding to
their finite values $\hbar \textbf{k}_f$ and $E_f = (\hbar k_f)^2/2 m_n$. Function $\widehat{U}(k)$ is
the Fourier transform of the neutron-helium atom interaction. In the range of large energies and momenta
transfer
considered in \cite{Dubnagroup01}, one can make use in (\ref{1-HP}) the \textit{pseudo-potential}
approximation with the corresponding scattering length $a$, see \cite{Dubnagroup1}(2). Then (\ref{1-HP})
can be rewritten as
\begin{equation}\label{1-HPb}
\frac{1}{N} \, \frac{d^2 \sigma}{d\Omega dE_f} = \frac{\sigma_b}{8 \pi^2 \hbar}
\left(1 - \frac{\hbar \omega}{E_i}\right)^{1/2}{S(k,\omega)} \, , \quad E_f = E_i - \hbar \omega \, ,
\end{equation}
where $\sigma_b := (1 + m_n/M_{He})^2 \, 4\pi a^2$ is the {bound-helium-atom} cross section.

Taking into account that $S(k,\omega)$ in (\ref{1-HP}) (or in (\ref{1-HPb})) is the Fourier transform of
\textit{density-density} correlation function (\cite{vanHove}, \cite{AksBal2023} \S4.2),
Hohenberg and Platzman \cite{HohenbergPlatzman} \textit{argued} that for \textit{high-energy} incident
neutrons and \textit{deep-inelastic} scattering (\textit{high} transfer of energy and momentum) the
{van Hove} factor gets the form (\textit{Impulse Approximation}):
\begin{equation}\label{2-HP}
S_{\beta,\mu }^{IA}(k,\omega)= \lim_{V\rightarrow\infty}\frac{1}{V}\sum_{\textbf{q}\in\Lambda^{\ast}}
n_q^{He}(\beta,\mu) \ \delta \, ( \hbar \omega - \frac{\hbar^2}{2 m} (\textbf{k}^2 +
2 \, \textbf{k}\cdot\textbf{q}) \,) \, ,
\end{equation}
where  $n_q^{He}(\beta,\mu ) := \left\langle N_{q}\right\rangle_{H_{\Lambda }^{He}}(\beta,\mu )$,
are mean-values of occupation numbers in modes $\textbf{q}\in\Lambda^{\ast}$ for atoms
of liquid helium $^{4}\rm{He}$, cf. (\ref{q-dens}) for the {ideal} Bose-gas.

Recall that the {high-energy} neutrons \textit{deep-inelastic} scattering (also called the
\textit{neutron Compton scattering} \cite{AksBal2023} \S5.4, 5.4.2) is such that
during the short time interval $t_S$ of a collision, the force of impact is much larger than all the
other forces. Therefore, counted in this period (the \textit{short-time scattering} approximation) we can
consider the other forces to be negligible. Then a \textit{noncoherent} scattering will be essentially
defined by \textit{momentum distribution} of individual targets (atoms of $^{4}\rm{He}$). This is
expressed by the formula (\ref{2-HP}). For that reason, the accuracy of the \textit{Impulse Approximation}
gets better
for growing values of $(k,\omega)$ (that is, for shorter $t_S$) and gives a direct information about
momenta of helium atoms. This would manifest as a \textit{two-component} structure of the van Hove factor,
which is the sum of a "narrow pick" corresponding to the scattering on a \textit{condensate} and a
"wide background" corresponding to the scattering on moving helium atoms \textit{out} of the {condensate}.

Since values of momentum and energy and transfers $(k, \hbar \omega)$ are bounded, one has to take into
account the impact of \textit{finite-state} interaction of the helium atom with environment, that
evidently breaks the kinematic relations in (\ref{2-HP}). As a consequence, the formula (\ref{2-HP})
needs corrections, see discussion in \cite{AksBal2023} \S5.4. On that account, it is instructive to
check the {two-component} \textit{ansatz} by applying (\ref{2-HP}) for the
\textit{ideal} Bose-gas, where there is no \textit{finite-states} interactions at all. Therefore, we
insert into formula (\ref{2-HP}) the ideal Bose-gas density distribution
$n_q (\beta,\mu) = \left(e^{\beta\, (\varepsilon_q - \mu)}- 1\right)^{-1}$, see (\ref{q-dens}).

For a given density of bosons: $\rho > 0$, on account of (\ref{2-HP}) and (\ref{5-g-c-densV})
one obtains for chemical potentials $\{\mu_\Lambda (\beta,\rho)\}_{\Lambda}$ that
\begin{equation}\label{3-HP}
S_{\beta,\mu_\Lambda (\beta,\rho)}(k,\omega)=
\lim_{V\rightarrow\infty}\frac{1}{V}\sum_{\textbf{q}\in\Lambda^{\ast}}
n_q(\beta,\mu_\Lambda (\beta,\rho)) \ \delta \, ( \hbar \omega - \frac{\hbar^2}{2 m} (\textbf{k}^2 +
2 \, \textbf{k}\cdot\textbf{q}) \,) \, .
\end{equation}
Then for $\rho > \rho_c(\beta)$, that is, in the presence of BEC, by virtue of  (\ref{3-g-c-densV}) and
(\ref{7-g-c-densV}) - (\ref{9-g-c-densV}) for $\lim_{V\rightarrow\infty} \mu_\Lambda (\beta,\rho)=0$,
we deduce from (\ref{3-HP}) for \textit{two} components of the van Hove dynamic structure form-factor
the following representation:
\begin{equation}\label{4-HP}
S_{\beta,\mu =0}(k,\omega)= \rho_0(\beta) \, \delta \, ( \hbar \omega - \frac{\hbar^2}{2 m} \textbf{k}^2)
+ \frac{m}{(2\pi\, \hbar)^2 k}\int_{\Delta (k,\omega)}^{\infty} dq \, {q} \,
\left(e^{\beta\, \varepsilon_q}- 1\right)^{-1} , \ \Delta (k,\omega) :=
\frac{|\hbar \omega - \varepsilon_k|\, m}{\hbar^2 k} \, .
\end{equation}

So, as we discussed above, for the ideal Bose-gas without the
\textit{final-state} atom interactions the neutrons scattering from "condensed" atoms yields in
dynamic structure function $S_{\beta,\mu =0}(k,\omega)$ (\ref{4-HP}) the $\hbar \omega$-dependent
\textit{narrow} "$\delta$-function" contribution supported on the \textit{one-particle} spectrum:
$\varepsilon_k = {(\hbar\textbf{k}) ^2}/{2 m}$,
see the first component in the right-hand side of equation (\ref{4-HP}). Interactions of the recoiled 
atoms
in the \textit{final-state} is the reason of broadening of this sharp pick.

The \textit{wider} last term (the second component) in the right-hand side of equation (\ref{4-HP}) is
a large pick as a function of transferred energy $\hbar \omega$ with maximum again at $\varepsilon_k$.
It is a result of the \textit{Doppler} broadening of the transferred energy owing to the neutron
scattering on the moving "non-condensed" atoms.

\subsection{Description of experiment and results (JINR 1975)}\label{subsection 2.3}

\subsubsection{Description of experiment}\label{subsubsection 2.3.1}

The experiment in \cite{Dubnagroup01} and \cite{Dubnagroup1} was carried out on the {IBR-30} pulsed
reactor in buster regime, using a {DIN-1M} spectrometer.
A monochromatic beam of neutrons with energy $E_i = 189,4 \, MeV$ was analysed, after scattering on the
sample at angle $\theta = 122,62^{\, \circ}$, by the \textit{time-of-flight} method between the sample and
the detector. Values of the transferred momentum that were most favorable for the experiment were chosen
in the range $k \sim (13\, \mathring{A}^{-1}-15 \, \mathring{A}^{-1})$ . The lower bound is determined
(for \textit{deep-inelastic scattering}) by the closeness of the approach to \textit{unity} of the ratio
of the energy transferred in the neutron scattering to the energy of the free helium atom that corresponds
to the momentum $\hbar k$: $\varepsilon_{He}(k) = (\hbar k)^2/2M_{He}$. This ratio was of the order of
$0,96$ at $k = 14,1 \, \mathring{A}^{-1}$ in our experiment \cite{Dubnagroup0}. This
interval is limited above by the resolving power of the spectrometer. In this experiment, the width of the
resolution function in the region of the "helium" peak was equal to the value $\sim9$ MeV.

We did not strive for the limiting parameters of the resolution function, since the condensate part
of the "helium" peak is broadened by the interaction in the \textit{final state} to the extent that it
is not possible to separate it in explicit form. Therefore, attention was concentrated on lowering the
statistical error and obtaining the highest possible accuracy in measurement of the shape of the "helium"
peak.

Over the time of the neutron-scattering experiment at $1,2 \,K$, the integrated count in the
"helium" peak amounted to $\sim2 \times 10^5$ pulses. The experimental spectra of inelastic neutron
scattering by liquid helium at temperatures of $1,2 \,K$ and $4,2 \,K$ were measured.
The width of the time spectrum analyser channel was $8 \,\mu$sec. The total time of
measurement amounted to $240$ hours. The measurements at the different temperatures were not normalised.
The energy resolution,
was determined with the help of a \textit{vanadium} sample and converted for the inelastic-scattering
region.

The \textit{background} was measured during evacuation of the helium vapor over the liquid at the bottom
of the cryostat. The center of the elastic peak is located in the $172$-nd channel. The relatively large
scatter in the results at the wings of the "helium " peak is explained by the fact that the background
was measured over times \textit{less} than the effect and was not smoothed but was calculated from
the channel. The experimental results were corrected for the effectiveness of the detector.

\subsubsection{Observation of Bose-Einstein condensate in the superfluid He II
\label{subsubsection 2.3.2}}

A numerical analysis of the experimental data was carried out on the basis of the regularised
iterative process of Gauss-Newton \cite{Alex1, Alex2} (library program C\^{O}MPIL, C-401, Dubna).

As a result of this analysis, it has been established that the model with \textit{two} \textit{Gaussian}
components (cf.(\ref{4-HP})) for the double-differential cross section of deep-inelastic high-energy
neutron scattering by $^{4}\rm{He}$ describes the experimental data sufficiently well. It was also
established that complication of the two-component approximating model by the addition of more
\textit{trial} Gaussians does not improve it.

Further, by comparison of quantities of the $\chi^2$-criterion for the two temperatures (below and above
$\lambda$-point), we verify that the model with \textit{two} Gaussians is \textit{better} from
the viewpoint of the statistical criteria at $T_1 = 1,2 \,K$
whereas the model with \textit{one} Gaussian is \textit{better} at $T_2 = 4,2 \,K$.

Finally, based on this double-Gaussian observation, we obtain the estimate $(3,6 \pm 1,4)\%$ for 
the value
of the \textit{fraction} $\rho_0/\rho$ for the \textit{Bose-Einstein condensate} 
$\rho_0$ at $T_1 = 1,2 \,K$.
Our analysis was also provided an evidence that to \textit{lower} the impact of the interaction of
the helium atoms in the
\textit{final state} (thus to improuve a reliability of the estimate of the {fraction} $\rho_0/\rho$),
the Bose condensate must be studied at \textit{higher} energies of the incident neutrons
together with improvement of the apparatus \textit{resolution function}.

For further discussion of the last subtle point concerning a balance between \textit{Doppler broadening}
and the \textit{width} of the resolution function for \textit{increasing} energy of the incident neutrons,
we refer to a very complete review \cite{Kozlov}, Ch.3, Sec.3.1.

\subsection{Observation of temperature dependence of Bose-Einstein condensate in the liquid
$^{4}\rm{He}$}\label{subsection 2.4}

At that time the Laboratory of Neutron Physics of JINR became also a pioneer in observation of
\textit{temperature dependence} of the Bose-Einstein condensate in liquid $^{4}\rm{He}$, 
\cite{Dubnagroup2}.
Besides, in this way one can revise the long-time \textit{open question} about {emergence} of the
condensate in liquid $^{4}\rm{He}$ together (\textit{simultaneously} !) with the superfluidity.

Experiment in \cite{Dubnagroup2} was carried out using the \textit{time-of-flight}
method with a DIN-1M spectrometer in the booster regime of the IBR-30 reactor, which is similar to
that in {subsubsection} \ref{subsubsection 2.3.1}.
The spectra of the neutrons scattered by the liquid helium $^{4}\rm{He}$ were measured simultaneously
at three scattering angles: $\theta = 122,6^{\, \circ}, 109,5^{\, \circ}$ and $96,5^{\, \circ}$ at for
initial neutron energy $E_i = 190$ MeV.
The aim was to analyse the \textit{temperature dependence} of the relative density of the Bose-Einstein
condensate in liquid $^{4}\rm{He}$ by studying the spectra of deep-inelastic neutron scattering 
at momentum
transfers corresponding to $k = (12 \, \mathring{A}^{-1}- 14 \, \mathring{A}^{-1})$ and temperatures
$T = (1,2 \,K - 4,2 \,K)$.

Note that there it was used in the measurement a cryostat with $^{4}\rm{He}$ vapor pumped on.
The helium temperature was determined by measuring the vapor pressure over the liquid
and with the aid of a carbon resistor. The temperatures
in the intervals $1,2 \,K - 1,8 \,K$ and $1,95 \,K - 4,2 \,K$ were
maintained with accuracy to $\sim 0.02 \,K$ and $\sim 0.01 \,K$, respectively.\\
The backgrounds of the empty cryostat and of the cryostat filled with liquid helium were determined by
performing the measurements alternately using a cadmium shutter to block the neutron beam and without
the shutter.

The main results of the experiment \cite{Dubnagroup2} are the following (below we include some precisions
due to later experimental data from \cite{Kozlov}):\\
1. The relative density $\rho_{0}/\rho$ of the Bose-Einstein condensate was calculated similarly to
\cite{Dubnagroup01, Dubnagroup1}, by the method of \textit{two-Gaussian} resolution the
spectra of neutrons scattered by liquid helium $^{4}\rm{He}$ in the temperature
interval $T = (1,2 \,K - 1,8 \,K)$. \\
2. The Bose-Einstein condensate was observed for  $T < T_0$, whereas for  $T \geq T_0$, within the
limits of the accuracy of the experiment and of the mathematical reduction, \textit{no} Bose-Einstein
condensate was observed. The relative density $\rho_{0}/\rho$ of the Bose-Einstein condensate was
estimated using formula
\begin{equation}\label{5-HP}
\frac{\rho_0}{\rho} = \frac{S_{BC}}{S_{BC} + S_{SC}}\, .
\end{equation}
Here in the frame-work of the \textit{two-Gaussian} resolution the value of $S_{BC}$ is the \textit{area}
below the  spectrum of neutrons scattered on the Bose-Einstein condensate, thought $S_{SC}$ (in the
two-Gaussian resolution) is the \textit{area} below the spectrum, which is due to the scattering on
non-condensed atoms of the liquid helium $^{4}\rm{He}$. Then temperature dependence of the relative
density fits into the form \cite{Dubnagroup2}:
\begin{equation}\label{6-HP}
\frac{\rho_0}{\rho} = \xi_0 \left(1 - \left(\frac{T}{T_0}\right)^m\right) \, ,
\quad T \leq T_0 = (2,22 \pm 0,05) \,K \, ,
\end{equation}
where
\begin{equation}\label{7-HP}
\xi_0 = (7 \pm 0,5)\% \, \quad \rm{and} \quad m = 9\pm4 \, .
\end{equation}
The analysis is essentially true in the \textit{vicinity} of the point $T_0$ and appeals to more
accuracy.\\
3. The Bose-condensation temperature $T_0$ (\ref{6-HP}) within the limits of the \textit{accuracy}
of the experiment is very close to the temperature $T_\lambda = 2,172 \,K$ ($\lambda$-point) 
of transition
the liquid helium $^{4}\rm{He}$ into {He II}, that is, into \textit{superfluid} phase
for decreasing temperature.\\
4. Character of the temperature dependence of the Bose-Einstein condensate relative density (\ref{6-HP})
is \textit{similar} to the temperature dependence of the relative density
\begin{equation}\label{8-HP}
\frac{\rho_s}{\rho} = 1 - \left(\frac{T}{T_\lambda}\right)^{5,6} \, ,
\quad T \leq T_\lambda = 2,172 \,K \, ,
\end{equation}
for \textit{superfluid} component $\rho_s$, see \cite{Andron46}, \cite{Khalat71} and
\cite{Huang1} Ch.13, Sec.13.2, whereas for the ideal Bose-gas index $m = 3/2$, 
\cite{Kvasnikov} Ch.III, \S2.

\vspace{0.2cm}

In conclusion we mention a radically different method \cite{ZagPri76} for studying the Bose-condensation
in liquid helium $^{4}\rm{He}$, see also  \cite{Kozlov} and \cite{Dubnagroup2}.
One of the possible way for measuring the Bose-Einstein condensate density and investigating the
connection between \textit{condensation} and \textit{superfluidity}, consists in performing experiments
with deep inelastic scattering of neutrons on liquid helium $^{4}\rm{He}$ with a small \textit{admixtures}
of the helium $^{3}\rm{He}$.

The presence of $^{3}\rm{He}$ with concentration $c>0$ shifts the \textit{superfluid transition}
(along the $\lambda$-line on the phase-diagram of the $^{3}\rm{He}$-$^{4}\rm{He}$ mixture) to lower
temperatures $T_\lambda (c>0) < T_\lambda$. For that reason $^{3}\rm{He}$ impurities are able to
destroy superfluidity and thus the Bose-Einstein condensate, see \cite{Eselson} and \cite{Betts}.
This observation and similarity between (\ref{6-HP}) and (\ref{8-HP}) would allow to establish a
\textit{direct} correlation between relative densities of Bose-condensation and superfluidity in He II.

To this aim a description of possible experiment has been presented in \cite{ZagPri76} for $c = 5\%$.
Then $T_\lambda (5\%) = 2 \,K$ and by (\ref{6-HP}) one obtains that for \textit{pure} system
at this temperature $({\rho_0}/{\rho})(c=0, T=2 \,K) \simeq 2\%$. If there is a
\textit{correlation} between superfluidity and Bose-condensation, then since for $c = 5\%$ one obtains
$({\rho_s}/{\rho})(c=5\%, T=2 \,K) =0$, one \textit{should} also observe the same value:
$({\rho_0}/{\rho})(c=5\%, T=2 \,K) =0$ in the deep-inelastic neutron scattering experiments.

Calculations in \cite{ZagPri76} showed that in view of the \textit{large} cross section
for neutron \textit{capture} by $^{3}\rm{He}$, to maintain the accuracy of observations, the
neutron flux density in this experiment must be \textit{larger} by approximately two orders of magnitude
than the existing flux. Another problem related to this phenomenon is a supplementary \textit{heating}
because of reaction:
\begin{equation}\label{9-HP}
n + \, ^{3}\rm{He} \ \rightarrow  \ ^{4}\rm{He} + (\sim 0,766 \, MeV) \, ,
\end{equation}
as well as  a (tiny) gradual \textit{decreasing} of the concentration $c$. The estimates in 
\cite{ZagPri76}
revealed that the problem of heating is soluble, but the first problem needs a new source of
neutrons and new spectrometer. 

For recent results on the properties of He II, including some experimental data about the 
Bose-Einstein condensate in liquid helium $^{4}\rm{He}$, see (a quite biased) review article 
\cite{Gly18}.

\vspace{-1.0 cm}

\begin{flushright}
\parbox{7.95cm}{``To construct a complete molecular theory of superfluidity it is
necessary to consider the liquid helium as being a system of interacting
atoms''\newline
\textbf{Nikola\"{\i} Nikolaevitch Bogoliubov}, \textit{Lectures on Quantum Statistics.}\newline

``Faced with this failure, theorists retreated into the corner of
low density gases with weak interaction''\newline
\textbf{Elliott H.Lieb}, \textit{The Bose Fluid.}}

\end{flushright}

\vspace{0.5cm}

\section{Condensation and Bogoliubov Theory of Superfluidity \label{section 3}}

After a guess about correlation between superfluidity and Bose-Einstein condensate in the liquid
$^{4}\rm{He}$ formulated by F. London in 1938, \cite{London} \S3, it was N.N. Bogoliubov
who proposed in 1947, \cite{Bogoliubov1}, \cite{Bogoliubov1bis}, \cite{Bogoliubov1bisbis}, an elegant
theory that \textit{links} the Bose-Einstein condensation with superfluidity. As a matter of fact,
F.London was trying to develop a "highly idealised model" for superfluidity of liquid
$^{4}\rm{He}$ by taking into account only the ideal Bose-gas with condensate and the quadratic
\textit{one-particle} spectrum: $\varepsilon_\textbf{k} = \hbar^2 \textbf{k}^2/(2m)$, see
\cite{London} \S3. However, according to the arguments based on the \textit{Landau criterion} of
superfluidity \cite{Landau1, Landau2}, the ideal Bose-gas with condensate can {not}
manifest the superfluidity ({zero viscosity}) because of these  \textit{low-lying} quadratic
one-particle excitations, cf. \cite{Landau1}.

In \cite{Bogoliubov1, Bogoliubov1bis} and \cite{Bogoliubov1bisbis} Bogoliubov declared that
for superfluidity it is necessary to consider the liquid helium as being a system with
\textit{interaction} and also with Bose-Einstein \textit{condensate} favouring a
\textit{collective} (instead of \textit{one-particle}) excitations of the condensed "helium jelly".
On that account, Bogoliubov selected as an essential \textit{ansatz}
the interaction between \textit{condensate} and \textit{out-of-condensate} atoms,
which is such that it is \textit{weak} enough to preserve condensate in the zero-mode $\textbf{k}=0$.
This \textit{off-diagonal} interaction in the Bogoliubov Weakly Imperfect Bose-Gas (WIBG)
\cite{BruZagrebnov3} yields necessary modification of the low-lying spectrum and produces the
\textit{collective} Bogoliubov excitations. Since they satisfy the \textit{Landau criterion}
\cite{Landau1}, this \textit{ansatz} completes the Bogoliubov theory of superfluidity
\cite{Bogoliubov1, Bogoliubov1bis}.

It is worth mentioning here that particle interaction may (in turn) to modify the \textit{nature} of the
Bose-Einstein condensate to a \textit{generalised} or \textit{non-conventional} one. To this end, see
subsections: {{Generalised (Bose-Einstein) condensation \textit{\`{a} la} van den Berg-Lewis-Pul\`{e}}}
\ref{subsubsection 1.2.2} and {{non-conventional (Bose-Einstein) condensation}} \ref{subsubsection 1.2.3}.
This, for example, would have impact on the properties of the WIBG. For further details about condensate
in the Bogoliubov model of WIBG and superfluidity, see \cite{ZagBru01} Sections 5 and 6.

\vspace{0.5cm}

\noindent
\textbf{P.S.}

\vspace{0.2cm}

\noindent
I would like to note that excellent review article \cite{VYIC2013} covers a \textit{point}, which
is deliberately missed in the present {double-jubilee} message dedicated \textit{uniquely}
to {discovery} of the Bose-Einstein condensation (1925) and to {observation} of the
Bose-Einstein condensation in the superfluid helium $^{4}\rm{He}$ (JINR 1975).
This \textit{point} involves  enormous literature about a special kind of condensation of dilute
ultracold atomic Bose-gases in traps, which, for example, inherits in \cite{BZ2010} some ideas
of Subsection \ref{subsection 1.2} related to different condensations. Curious readers will find
a lot of information about this missed \textit{point} in paper \cite{VYIC2013}.

Another important missing \textit{point} is mentioned in the \textit{epigraph} by {Immanuel Kant},
which is quoted before Section \ref{section 1}. It concerns the mathematical \textit{status} of the
Bose-Einstein condensation concept. For the mathematically-mind readers it would instructive to consult
first an excellent, short and comprehensive survey \cite{LSSY2005} about a variety of mathematically 
rigorous
results, including also those on the \textit{ultracold} atomic Bose-gases in traps. Besides, there is a
kind of mathematical results, for example \cite{LewisPule1}, or those collected in the books
\cite{Verbeure}, \cite{BratteliRobinson}, which study the Bose-Einstein condensation in the framework
of the abstract mathematical approach to the Quantum Statistical Mechanics.

\vspace{2cm}

\noindent
\textbf{Acknowledgements}\\

\noindent
First of all I am thankful to Irina Aref'eva and Igor Volovich for discussions of the topics
and of the relevance of this article. They convinced me that
this \textit{anniversary recall} about jubilee of the Bose-Einstein condensation (1925), as well as
about its observation in deep-inelastic neutron scattering experiments involved the superfluid helium
$^{4}\rm{He}$ (JINR-Dubna 1975), may be of interest to readers.

During my recent visit JINR-BLTP (May-July 2025) I had useful discussions with Alexander Povolotsky
on different aspects of the Bose-Einstein condensation. My gratitude is owed to him for attention
and hospitality. 

I am also indebted to Viktor Aksenov, Zbigniew Strycharski and Efim Dynin
for useful discussions, suggestions and for help with references as well as with collection
of some rare publications.

Finally, I am very thankful to Joseph Pul\'{e}, one of the authors of the concept of \textit{generalised}
Bose-Einstein condensation, for his useful remarks and a number of important suggestions.

\vspace{1cm}

\newpage

\end{document}